\documentclass[conference,compsoc]{IEEEtran}
\IEEEoverridecommandlockouts

\usepackage{cite}
\usepackage{amsmath,amssymb,amsfonts}
\usepackage{algorithmic}
\usepackage[linesnumbered,ruled,vlined]{algorithm2e}
\usepackage{graphicx}
\usepackage{subfigure}
\usepackage{textcomp}
\usepackage{xcolor}
\usepackage{booktabs}       
\usepackage{multirow}       
\usepackage{makecell}
\usepackage{subcaption}  
\usepackage{diagbox}
\usepackage{threeparttable}

\def\BibTeX{{\rm B\kern-.05em{\sc i\kern-.025em b}\kern-.08em
    T\kern-.1667em\lower.7ex\hbox{E}\kern-.125emX}}
\begin{document}

\title{FastFHE: Packing-Scalable and Depthwise-Separable CNN Inference Over FHE\\
}

\author{%
Wenbo Song$^{1,3}$ \enspace Xinxin Fan$^{1,2}$ \enspace Quanliang Jing$^{1}$ \enspace Shaoye Luo$^{1,2}$ \enspace Wenqi Wei$^3$ \enspace Chi Lin$^4$  
\enspace Yunfeng Lu$^5$ \enspace Ling Liu$^6$ \\[-0.2em]
$^1$Institute of Computing Technology, CAS \quad $^2$UCAS \quad $^3$Fordham University 
\quad $^4$Dalian University of Technology \\[-0.2em] 
\quad $^5$Beihang University
\quad $^6$Georgia Institute of Technology \\[-0.3em] 
\texttt{songwenbo@mail.dlut.edu.cn} \enspace \texttt{\{fanxinxin,jingquanliang, luoshaoye\}@ict.ac.cn}\\[-0.3em]
\texttt{c.lin@dlut.edu.cn} \enspace \texttt{wenqiwei@fordham.edu} 
\enspace \texttt{lyf@buaa.edu.cn} \enspace \texttt{lingliu@cc.gatech.edu}
}

\maketitle

\begin{abstract}
The deep learning (DL) has been penetrating daily life in many domains, how to keep the DL model inference secure and sample privacy in an encrypted environment has become an urgent and increasingly important issue for various security-critical applications. To date, several approaches have been proposed based on the Residue Number System variant of the Cheon-Kim-Kim-Song (RNS-CKKS) scheme. However, they all suffer from high latency, which severely limits the applications in real-world tasks. Currently, the research on encrypted inference in deep CNNs confronts three main bottlenecks: i) the time and storage costs of convolution calculation; ii) the time overhead of huge bootstrapping operations; and iii) the consumption of circuit multiplication depth. 

Towards these three challenges, we in this paper propose an efficient and effective mechanism FastFHE to accelerate the model inference while simultaneously retaining high inference accuracy over fully homomorphic encryption. Concretely, our work elaborates four unique novelties. First, we propose a new scalable ciphertext data-packing scheme to save the time and storage consumptions. Second, we work out a depthwise-separable convolution fashion to degrade the computation load of convolution calculation. Third, we figure out a BN dot-product fusion matrix to merge the ciphertext convolutional layer with the batch-normalization layer without incurring extra multiplicative depth. Last but not least, we adopt the low-degree Legendre polynomial to approximate the nonlinear smooth activation function SiLU under the guarantee of tiny accuracy error before and after encrypted inference. Finally, we execute multi-facet experiments to verify the efficiency and effectiveness of our proposed approach, and the experimental results show that, at the standard 128-bit security level, our FastFHE enables the popular ResNet20 to achieve a 2.41$\times$ reduction in inference latency and a 2.38$\times$ descent in amortized runtime (using 30 threads) compared to the state-of-the-art methods. Moreover, the performance on several other representative deep CNN architectures, such as ResNet32, ResNet44 and VGG11, also exhibits remarkable advantages of our work in both inference overhead and accuracy.
\end{abstract}


\section{Introduction}
In today's digital era, spurred by the rapid advancement of artificial intelligence in recent years, machine learning has achieved significant breakthroughs in both software and hardware. As a result, it has gradually become the future direction of development across a wide range of industries. As data volume grows and hardware computing power continues to arise, Machine Learning as a Service (MLaaS) has become a preferred choice for many professionals \cite{HuangXu25}. MLaaS is a cloud-based service whose primary goal is to lower the entry barriers to machine learning, enabling users from various fields to focus on their business demand without worrying about hardware resources, algorithm implementation, or deployment details \cite{dosovitskiy2020image, fawzi2022discovering}. However, MLaaS not only encounters conventional adversarial attacks \cite{WangFJTB21, WangFJSLB22, LuoFan25}, but it also suffers from privacy-specific breaches both on sample data and model information,
such as membership inference attack \cite{LiuHan2024, JiachengLi2024, YuHe2024, ZitaoChen2025}, model inversion \cite{ShanghaoShi2025}, and model extraction \cite{PeizhuoLv2024, MinxueTang2024, TusharNayan2024}. One main concern of the clients is the sample data leakage, that is, uploading sensitive data to the cloud server may bring in huge risks of information leakage and privacy breach, which profoundly prompts researchers to look into the issue of privacy-preserving machine learning (PPML) in real-world applications.  

PPML is a technical paradigm designed to accomplish machine learning tasks while ensuring data privacy. It addresses the issue of sensitive-data leakage in the pipeline of machine learning, such as medical records, financial transactions, personal behavioral data, etc. Surrounding the core PPML techniques, the fully homomorphic encryption (FHE), which keeps the data encrypted throughout the entire computation process, provides the highest level of privacy protection in theory. It has increasingly become the choice for many researchers to execute PPML.

\begin{figure}[tbhp] 
\centering
\includegraphics[width=1.0\linewidth]{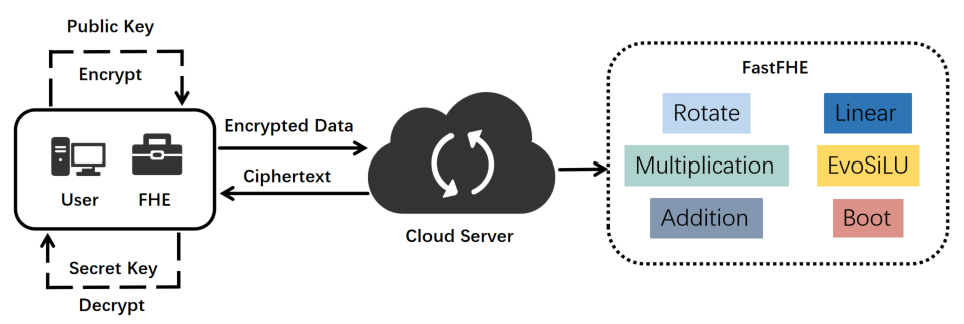}
\caption{MLaaS over FHE. The dashed box highlights the homomorphic operations in our FastFHE.}
\label{fig:MLaaS}
\end{figure}

As the achievement of deep neural networks, secure inference for deep-learning models in a FHE environment is becoming a promising privacy-preserving MLaaS solution at present. As we know, FHE supports performing arbitrary homomorphic additions and homomorphic multiplications on ciphertext without the risk of decryption errors \cite{gentry2009fully, cheon2018bootstrapping}. Fig. \ref{fig:MLaaS} briefly illustrates the process of secure deep CNN inference over FHE. That is, a user encrypts private data using the FHE public key and sends the encrypted data to the server in the cloud. The server performs deep CNN inference on the encrypted data and returns the encrypted inference result to the user/client, who would decrypt it with the owned private key to obtain the final plaintext outcome. Throughout this pipeline, the user's private data remains encrypted during the entire inference process, ensuring the server side cannot access any data information.

Currently, scholars have predominantly focused on constructing deep CNNs over FHE, for example, Gilad-Bachrach et al. \cite{gilad2016cryptonets} and Badawi \cite{al2020towards} had successfully deployed neural networks in the encrypted environment. Nevertheless, the practicality is severely constrained by the shallow network depth and the absence of non-linear activation functions, this is due to the available multiplicative depth is limited in leveled homomorphic encryption (Leveled-HE). Subsequently, the FHE is employed to build deeper neural networks, and the bootstrapping operation \cite{cheon2018bootstrapping} becomes necessary, this is because it can refresh both the noise level and the multiplicative depth of ciphertext computation, which enables to effectively transform the Leveled-HE into FHE and make it possible to construct arbitrarily deep neural networks. In the reference \cite{lee2022privacy}, the bootstrapping operation was used for the first time to realize a deep neural networks over FHE. However, the approach employed a high-degree polynomial to approximate the activation function ReLU, which consumed considerable multiplicative depth and required huge bootstrapping operations, ultimately resulting in a prolonged inference time.

As well known, the information propagation undergoes a long path along layers in the deep CNNs, thus, the operating in encrypted data over FHE accordingly costs a substantial time, a regular and valid solution is to optimize the procedure of convolution computation via an appropriate data-packing scheme, with the purpose of reducing the overall overhead. For instance, GAZELLE \cite{juvekar2018gazelle} offers an efficient FHE-matched convolution algorithm that significantly lowers the number of homomorphic operations needed for traditional convolution in the ciphertext domain. However, in deeper neural networks, the increasing number of input/output channels still leads to high overhead. Furtherly, building on GAZELLE \cite{juvekar2018gazelle}, Lee \cite{lee2022low} and Lee \cite{lee2022privacy} propose more efficient convolution algorithms and succeed in implementing deeper neural networks. Nonetheless, due to the high computation cost in convolution and the substantial multiplicative depth required for the activation function ReLU, their approaches still demand more overhead.

In addition, for the commonly-used nonlinear activation function, e.g. ReLU, given the FHE only supports homomorphic addition and multiplication operations, hence, the approximation computation on activation functions is usually to use high-degree polynomial. Nevertheless, this causes another problem, i.e., such high-degree polynomial consumes a large amount of multiplicative depth, leading to more frequent bootstrapping operations and significantly prolonging overhead. For instance, Lee et al.\cite{lee2022privacy} used degree-5/15/27 polynomials to approximate the activation function, resulting in a large number of bootstrapping operations. AESPA \cite{park2022aespa} addresses this challenge by proposing a low-degree polynomial to reduce the overhead of approximately computing ReLU. Despite the overhead is reduced, the inference accuracy declines dramatically. Similarly, Ao and Boddeti \cite{ao2024autofhe} adopts a multi-degree polynomial associated with the layered structure, although the optimization on the placement of bootstrapping enables to minimize the amount of bootstrapping operations during convolution, it cannot benefit both efficiency and inference accuracy, still requiring considerable time cost.

To address the highly time-consuming problems while simultaneously retaining the inference accuracy, we propose a ciphertext data packing-scalable and block-ciphertext depthwise-separable CNN inference mechanism FastFHE. Briefly, our work focuses on the core-contents of PPML, i.e. data-packing scheme, multi-channel ciphertext convolution, and approximation expression of activation function. Our solution can achieve a fast and accurate inference for deep CNNs over FHE. Four main contributions are involved:

\begin{itemize}
\item {\textbf{Block-Ciphertext Depthwise-Separable Multi-Channel Convolution.}} Traditional convolution under ciphertext grows increasingly time-consuming as the number of output channels rises. 
Towards this problem, our work proposes a novel scalable data-packing scheme to remarkably reduce the convolution overhead by approximately a factor of $2/(K^2+1)$ ($K$ is the kernel size) through optimizing the usage of rotation keys and rearranging the accumulation of feature values. The more the channels, the more pronounced the reduction in overall cost.
\item{\textbf{Low-Degree Legendre Polynomial Approximation of Activation Function.}} In PPML environment, the currently existing methods usually approximate the commonly-used activation function ReLU using either high-degree Chebyshev polynomial, which requires excessive multiplicative depth, or low-degree Hermite polynomial, which compromises model inference accuracy. We think it is the ReLU's sharp corner (non-smooth) that makes it difficult to achieve good approximation with low-degree polynomial. Towards this problem, we adopt a low-degree Legendre polynomial to approximate other smooth activation function-SiLU, to gain a balance between acceptable multiplicative depth consumption and highly-accurate approximation.
\item{\textbf{Fusing Ciphertext Convolutional Layer with Batch-Normalization Layer.}} Without increasing extra multiplicative depth consumption, we propose a new architecture ConvBN to fuse the ciphertext convolutional layer and batch-normalization layer, which not only reduces the overhead of linear-layer multiplicative depth to a large extent, but also notably descends the overhead of activation-layer multiplicative depth by low-degree polynomial.
\item{\textbf{Multi-Facet Experiment Validations on Efficiency and Inference Accuracy.}} We implement extensive experiments on ResNet20/32/44 \cite{KaimingHe2016} and VGG11 \cite{KarenSimonyan2015} using RNS-CKKS scheme \cite{cheon2018full, cheon2017homomorphic}. For ResNet20, the results show the inference latency is 763s with an amortized execution time of 26s in a 30-thread environment, and simultaneously the accuracy remains high after encrypted inference. Compared to the state-of-the-art MPCNN \cite{lee2022low}, AESPA \cite{park2022aespa}, AutoFHE \cite{ao2024autofhe}, our FastFHE reduces the latency by 3.26$\times$, 2.51$\times$, 2.41$\times$, and lowers the amortized time by 3.19$\times$, 2.46$\times$, 2.38$\times$, respectively.
\end{itemize}

\section{Background and Problem Statement}
\subsection{Homomorphic-Operation Consumption}

FHE is an ideal lattice-based encryption mechanism and characterized by the ability to perform computation on encrypted data \cite{gentry2009fully}, the core of which lies in allowing to perform addition and multiplication operations on ciphertext, achieving data privacy-preserving. Given the advantage that FHE enables to continuously execute computation on encrypted data until the decryption is required, it has gained more and more attention in the field of PPML. RNS-CKKS \cite{cheon2018full, cheon2017homomorphic} is recently proposed to compute floating-point data in FHE field. It integrates CKKS encryption scheme with Residue Number System (RNS), and significantly improves the efficiency of parallel operations. As well known, the CKKS is a homomorphic encryption scheme designed for floating-point computation, it encodes multiple data elements into a polynomial form within a single ciphertext and supports homomorphic addition and multiplication. Meanwhile, through splitting data into multiple parts, RNS effectively reduces the complexity of homomorphic operations. 

In CKKS, a plaintext $m$ is first mapped to a polynomial in the ring
\begin{math}
	\mathbb{Z}_{q}[X] /\left(X^{D}+1\right)
\end{math}.
Then, it is combined with the public key 
\begin{math}
	p_{k}=(a, e)
\end{math}
to perform addition and multiplication operations, producing the ciphertext 
\begin{math}
	c=a\cdot m+e
\end{math}.
Here, {\itshape D} is the ring dimension, {\itshape a} and {\itshape e} are noise terms from the public key. The original polynomial is further split across several prime moduli, by which the parallel computation can be realized. The partial results from multi-modulus computation are merged to obtain the final outcome, in this way, the overall computation efficiency can be improved to a large extent. The decryption is achieved through key-switching procedure, and the ciphertext noise from the individual prime modulus is transformed into the original domain. For each two ciphertexts 
\begin{math}
	ct_{1}=(c_{10},c_{11}) 
\end{math}
and 
\begin{math}
	ct_{2}=(c_{20},c_{21}) 
\end{math},
the homomorphic addition and multiplication are defined as:
\begin{equation}
	ct_{add}=(c_{10}+c_{20},c_{11}+c_{21}) \bmod q,
\end{equation}
\begin{equation}
	ct_{mult}=(d_{0}+d_{1})+RLK(d_{2}) \bmod q,
\end{equation}
where 
\begin{math}
	d_{0}=(c_{10}\cdot c_{20}) \bmod q
\end{math},
\begin{math}
	d_{1}=(c_{10}\cdot c_{21}+c_{11}\cdot c_{20}) \bmod q
\end{math},
\begin{math}
	d_{2}=(c_{11}\cdot c_{21}) \bmod q
\end{math},
\begin{math}
	RLK(d_{2})
\end{math} 
represents the relinearization operation, which applies the relinearization key 
\begin{math}
	rlk
\end{math} 
to transform 
\begin{math}
	d_{2}
\end{math} 
into a linear form.

\begin{table}[tb]
\caption{Execution time for various homomorphic operations, wherein "Pt" and "Ct" denote plaintext-ciphertext and ciphertext-ciphertext, "CRot" denotes ciphertext rotation, "Add" and "Mul" indicate homomorphic addition and multiplication, "Boot" represents bootstrapping.}
\label{tab:homomorphic_ops}
\begin{tabular}{>{\centering\arraybackslash}p{1.0cm} >{\centering\arraybackslash}p{0.75cm} >{\centering\arraybackslash}p{0.75cm} >{\centering\arraybackslash}p{0.75cm} >{\centering\arraybackslash}p{0.75cm} >{\centering\arraybackslash}p{0.75cm} >{\centering\arraybackslash}p{0.75cm}}
\toprule
\textbf{Operation} & \textbf{AddPt} & \textbf{AddCt} & \textbf{MulPt} & \textbf{MulCt} & \textbf{CRot} & \textbf{Boot} \\
\midrule
Time(s) & 0.20s & 0.22s & 0.57s & 1.26s & 1.06s & 14.176s \\
\bottomrule
\end{tabular}
\end{table}

Another important concept in FHE is the multiplicative depth closely tied to the ring dimension $D$, i.e. the larger the dimension $D$, the greater the maximum multiplicative depth $L$. Each homomorphic multiplication consumes one level of multiplicative depth. Once all levels are used up, a bootstrapping operation \cite{cheon2018bootstrapping} is required to refresh the ciphertext's multiplicative depth, its validity and precision. Table \ref{tab:homomorphic_ops} shows our experimental results on the time costs of various homomorphic operations under RNS-CKKS scheme, it reveals the bootstrapping is significantly more time-consuming than the other homomorphic operations. Therefore, reducing bootstrapping as much as possible is critical for efficient encrypted computation.

\textbf{Tremendous Bootstrapping and Numerous Rotation operations Are Required in Convolution Calculation.} During the convolution calculation over FHE, the data arrangement in ciphertext slots is equivalently important, as it affects the number of rotation and bootstrapping operations. The rotation refers to cyclically shifting the encrypted data for the task execution, such as rearranging data, performing encryption computation using specific patterns. On the other hand, when many homomorphic multiplications are needed, each ciphertext may require bootstrapping to continuously implement encrypted-data computation, resulting in frequent bootstrapping operations. Therefore, reducing the number of ciphertexts is beneficial to the efficiency. Furthermore, if a single ciphertext contains a large amount of data, the complex calculation would necessitate numerous rotations, as shown in Table \ref{tab:homomorphic_ops}, it is also time-consuming as well. 

To sum up, upon the statement above, we know that, in order to improve efficiency, it is crucial to strike a balance between packing large amount of data into fewer ciphertexts and minimizing costly rotation and bootstrapping operations.

\begin{figure*}[bt]
	\centering
	\includegraphics[width=6.4in, height=0.8in]{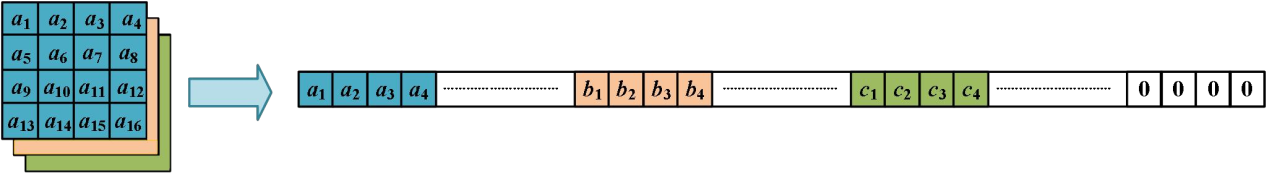}
	\caption{Packing image into ciphertext slots, wherein 
		\begin{math}
			a_{i}
		\end{math}, 
		\begin{math}
			b_{i}
		\end{math}, and 
		\begin{math}
			c_{i}
		\end{math} 
		denote feature values respectively from three channels. Any unused store slots are filled with zeros.
	}
	\label{fig:DataPacking}
\end{figure*}

\subsection{Existing Ciphertext-Convolution Solution}
For image-specific deep CNNs, such as object recognition, the convolution calculation plays a pivotal role in extracting local feature from the input data. For example, a 2D image is first converted into a three-dimensional tensor and then passed through convolutional filters/kernels. Let 
\begin{math}
	\left \{ c_{i},h_{i},w_{i} \right \} 
\end{math}
and 
\begin{math}
	\left \{ c_{o},h_{o},w_{o} \right \} 
\end{math} 
represent the input and output dimensions of a convolution layer, where {\itshape c} indicates the number of channels and {\itshape h}, {\itshape w} represent the height and width. The filter size in convolution is denoted as 
\begin{math}
	\left ( f\times f \right ) 
\end{math}.
Given ResNet \cite{KaimingHe2016} as a classic CNN architecture to achieve a deep learning in the true sense for the first time by introducing "deep residual connections", we subsequently discuss our work referring to ResNet as an example, and set a stride of $s=1, 2$ and padding to 1 in the convolutional layers.

Over FHE, the data in ciphertext slots is stored as one-dimensional vector. Hence, we must flatten the three-dimensional tensor data of the input image into a one-dimensional vector. Then, we can fill these ciphertext slots based on a specific data arrangement strategy as shown in Fig. \ref{fig:DataPacking}. Such a ciphertext data-packing method can straightly influence the computation efficiency of convolutional layer. As we know, the maximum size of feature map in ResNet20 is 
\begin{math}
	16\times 32\times 32=16384
\end{math},
thus, the number of store slots on-demand is set to
\begin{math}
	2^{14}=16384
\end{math}.

Towards the complex convolution calculation, Gazelle \cite{juvekar2018gazelle} proposed an efficient scheme SISO for the deep CNN inference over FHE, in which the 
\begin{math}
	h_{i}\times w_{i}  
\end{math} 
feature values are flattened and packed into a single ciphertext, along with the slot arrangement as shown in Fig. \ref{fig:DataPacking}. 

\textbf{Stationary Data-Packing and Idle Slots Raise Computation Complexity.} Investigating on the currently popular SISO scheme, we find the following observations and problems:  
i) since the input data is packed in a single ciphertext, the only way to execute convolution is to rotate the ciphertext. As a result, both the input ciphertext and convolution kernel need to be rotated according to different rotation indices, by which $f_h \times f_w$  rotated ciphertexts and kernel plaintexts are generated. The numbers of rotations and specific indices depend on two factors: the size of defined kernels and whether the feature values participate in the convolution operation; 
ii) each rotated ciphertext is then multiplied by its concordant rotated kernel plaintext, and the products from the $f_h \times f_w$  rotations are accumulated to produce the SISO convolution result;
iii) take CIFAR-10 dataset and ResNet\cite{KaimingHe2016} as an example, each image sample has 3 input channels (RGB), and the channel count gradually increases. Using SISO for the convolution operation would produce a large number of ciphertexts, leading to huge bootstrapping operations, which significantly extends the total computation time of deep CNNs over FHE;
and iv) on the other hand, when a large stride is set, e.g. $s=2$, the ciphertext obtained after convolution would contain more idle slots, which causes more influence on the efficiency. 

From these observations, we unveil two aspects that affect the computational complexity, i.e. the stationary block-packing scheme and idle slots incurred by larger stride. 

\subsection{Approximation of Activation Function}
Given that the ciphertext computation in FHE only supports homomorphic addition and multiplication, hence, we need elaborate an combination of such two operations to approximately compute the nonlinear activation function.

\textbf{Accuracy Degradation with Low-Degree Polynomial Approximation.} To achieve the approximation of nonlinear activation function, CryptoDL\cite{hesamifard2018privacy}, as one of the foundational works, employs polynomial to represent activation function in a homomorphic encryption (HE) environment. Since the bootstrapping is unavailable under HE, it cannot apply into deeper neural networks. Subsequently, some other works \cite{chabanne2017privacy, park2022aespa, ishiyama2020highly, obla2020effective, hesamifard2019deep} explore the low-degree polynomial for the activation-function approximation. One of which, AESPA \cite{park2022aespa} utilizes a degree-2 Hermite polynomial to approximate ReLU and gains a remarkable reduction on runtime. Nevertheless, the common limitation of such category of low-degree polynomial approximation methods lies in they cannot accurately represent the activation function, resulting in significant degradation on inference accuracy.

\textbf{Time-Consuming Computation with High-Degree Polynomial Approximation.} Take into consideration the accuracy, on the high-degree polynomial side, Lee et al. \cite{lee2022privacy} compare diverse configurations and find a combination of degree-5/15/27 minimax approximation polynomials can obtain an appropriate accuracy. This allows pre-trained models to be directly used, nonetheless, it still consumes substantial multiplicative depth stemming from numerous bootstrapping operations within the residual blocks. Towards this expensive cost problem, AutoFHE \cite{ao2024autofhe} recently frames this problem as a multi-objective optimization, and employs a layered and mixed-degree polynomial to seek balance between maximum accuracy and minimized number of bootstrapping operations, nevertheless, the large and sophisticated composition still demands extremely-high computational resources, e.g., at least 768GB of RAM, which makes it hard to deploy in regular environment without high-performance computing resources. 

Hence, a question emerges naturally, i.e. whether there exists a proper low-degree approximation polynomial to achieve a remarkable overhead reduction, and simultaneously maintain high inference accuracy. 

\subsection{Threat Model}
As stated in AESPA \cite{park2022aespa}, MPCNN \cite{lee2022low} and AutoFHE \cite{ao2024autofhe}, our proposed solution is applicable to their mentioned threat model as well. Specifically, we address the PPML for users' data in MLaaS scenarios, that is, when users rely on cloud server to execute CNN inference, they anticipate their sensitive data to be secure and not disclosed to the server. To this end, they usually encrypt their private data using a public key and upload both the ciphertext and public key to the server. The server then performs encrypted-data computation and inference on the ciphertext, and returns the computation result to the users. Then, they decrypt with their private key to obtain the final inference outcome. Throughout the entire process, the server only deals the ciphertext and cannot access to the private (plaintext) data. 

\section{FastFHE: An Efficient and Effective PPML}

\subsection{Feature Map-Aligned Block-Oriented Packing}
Diverse homomorphic operations have different computational overheads as reported in Table \ref{tab:homomorphic_ops}, from which we see the bootstrapping costs the most time. Thereby, we need to minimize the frequency of bootstrapping, and a direct and effective solution is to reduce the number of ciphertexts, especially when multiple channels exist in each layer. As we know,  the SISO \cite{juvekar2018gazelle} performs convolution calculation on a single-input/output channel, instead, our work aims to extend to multi-channel convolution and further reduce the cost of ciphertext rotation and multiplication operations. In detail, our approach employs a scalable $N$$\times$$N$ block-based data-packing scheme regarding the variational sizes of feature maps during the pipeline of convolution, by which all channels of feature maps can be packed into a single ciphertext. Of course, the ciphertext capacity, i.e. the number of slots, must accommodate the size of the largest feature map in the whole convolution pipeline. For example, upon the analysis on ResNet20, we know the largest feature map is: $16$$\times$32$\times$$32=16384$. As well known, for a CNN, the semantics-extracting process is usually to narrow down the size of feature map and extend the number of channels. Routinely, the stride may not be set too large, this is because large stride would cause much loss of local information, thus, the downsampling is usually to scale feature map to the half of original size, and correspondingly double the channel number. To comply with 2D matrix block, we defines the data-packing block parameter $N$ as:
\begin{equation}
	N=\left\{\begin{matrix}
		\frac{\left({F_{max}}\right)^{\frac{1}{2}}}{h_{i} } & s=1\\ \\  
		\frac{\left({F_{max}}\right)^{\frac{1}{2}}}{h_{o} } & s=2
	\end{matrix}\right.,
\end{equation}
where $F_{max}$ denotes the size of the largest feature map, e.g. 128 for ResNet20, $h_{i}$ and $h_{o}$ are input channel's height and output channel's height, and $s$ is the stride of convolution. 


\begin{figure} [t]
	\centering
	\subfigure[3$\times$32$\times$32 data packing]{
		\begin{minipage}[t]{0.22\textwidth}
			\centering
			\includegraphics[width=1.5in, height=1.0in]{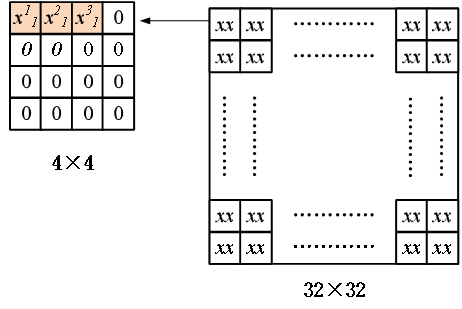}
			\label{Fig:sub1}
		\end{minipage}
	}
	\vspace{0.02cm}
	\subfigure[	16$\times$32$\times$32 data packing]{
		\begin{minipage}[t]{0.22\textwidth}
			\centering
			\includegraphics[width=1.5in, height=1.0in]{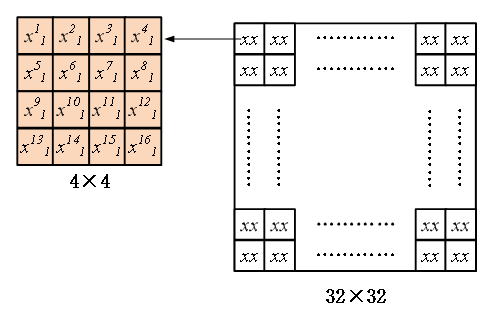}
			\label{Fig:sub2} 
		\end{minipage}
	}
	\vspace{0.02cm}
	\subfigure[32$\times$16$\times$16 data packing]{
		\begin{minipage}[t]{0.22\textwidth}
			\centering
			\includegraphics[width=1.6in, height=1.0in]{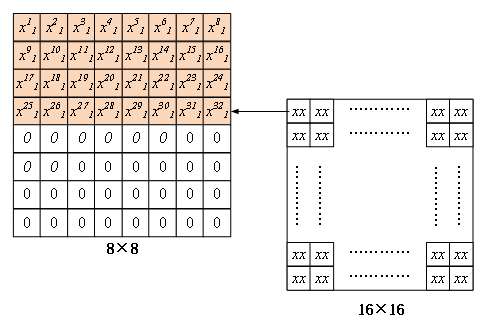}
			\label{Fig: PGD} 
		\end{minipage}
	}
    \vspace{0.02cm}
    \subfigure[64$\times$8$\times$8 data packing]{
    	\begin{minipage}[t]{0.22\textwidth}
    		\centering
    		\includegraphics[width=1.6in, height=1.0in]{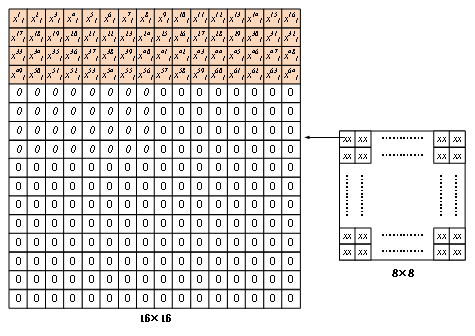}
    		\label{Fig: PGD} 
    	\end{minipage}
    }
	\caption{Scalable block-oriented packing scheme.}
	\label{fig:scalabelPacking}
\end{figure}	


We can comparatively analyze our method with the prevalent SISO scheme \cite{juvekar2018gazelle}, each position that stores a single element in SISO is currently treated as an $N$$\times$$N$ block. The size of block per row depends on the stride, and each $N$$\times$$N$ block stores the feature values from all channels. Once the number of channels is less than block size, the rest elements of the bock are padded with zeros as placeholders, keeping a complete square. Also take ResNet20 as an example, as shown in Fig. \ref{fig:scalabelPacking}, the ciphertext has $2^{14}$=16384 slots, which can be viewed as a 128$\times$128 square. This square is composed of multiple $N$$\times$$N$ blocks. For a 16$\times$32$\times$32 feature map, each 4$\times$4 block can store the 16-channel information at a single position. Similarly, Fig. \ref{fig:scalabelPacking}(b)-(d) give the continued downsampling layer by layer from feature-map size 32$\times$32 with channel number 16 to 16$\times$16 with channel number 32, and to 8$\times$8 with channel number 64. In this way, the multi-channel data is neatly packed into a single ciphertext. That is to say, our proposed multi-channel convolution can be carried out by manipulating just one ciphertext, which significantly reduces the computation and storage complexities, compared to the stationary SISO-packing scheme. Next, we design homomorphic convolution fashion to match our new data-packing scheme.


\subsection{Ciphertext Depthwise-Separable Convolution}
\textbf{Traditional Convolution Manipulation.} As stated previously, for each feature map during convolution, our method need have the capability to tackle $N$$\times$$N$ block ciphertext as input, and produce another $N$$\times$$N$ block ciphertext as output. Prior to clarifying our proposed block-ciphertext depthwise-separable convolution, we at first introduce the routine operation on ciphertext. Taking a 4$\times$4 feature map and 3$\times$3 convolution kernel as an example, we demonstrates the detailed steps for the traditional convolution manipulation in Fig. \ref{fig:ciphertextTraditionalConvo} in Appendix \ref{Sec:traditionCiphertextConvo}.

%
%
%

The traditional convolution scans the input data with a set of convolution kernels to extract local features. For a deep CNN, the beginning convolutional layers can capture shallow features, such as edges, textures and colors, while the deep layers enable to extract the complex and abstract feature. Therefore, in such a deep CNN, as the number of kernels or input channels grows, the computation load will become enormous, or even the bottleneck over FHE.

\begin{figure*} [hbtp]
	\centering
	\subfigure[Traditional convolution]{
		\begin{minipage}[t]{0.35\linewidth}
			\centering
			\includegraphics[width=2.0in, height=1.2in]{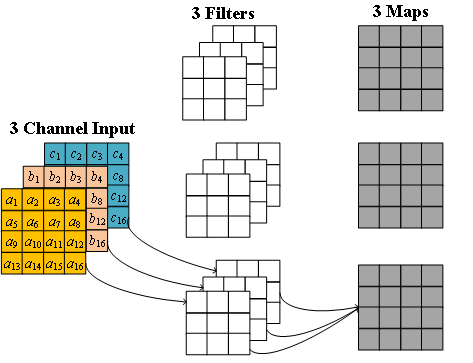}
			\label{Fig:TraditionalConv}
		\end{minipage}
	}
	\vspace{0.02cm}
	\subfigure[Depthwise-separable convolution]{
		\begin{minipage}[t]{0.35\linewidth}
			\centering
			\includegraphics[width=2.0in, height=1.2in]{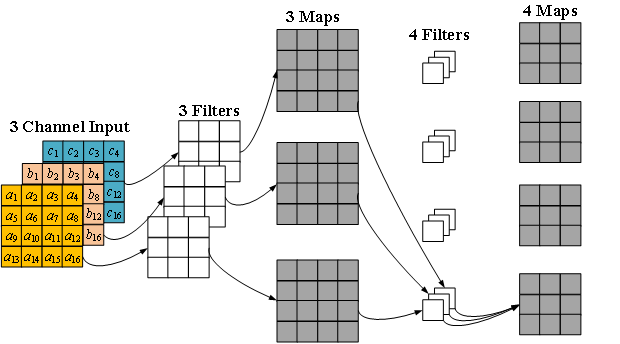}
			\label{Fig:DepthSepaConv} 
		\end{minipage}
	}
	\caption{Architectures of traditional convolution vs. depthwise-separable convolution.}
	\label{fig:traditionalDepthSeparableCompare}
\end{figure*}

\textbf{Block-Ciphertext Depthwise-Separable Convolution Manipulation.} The traditional convolution slides the kernel over the input image, implementing an elementwise multiplication and summation at each position to produce the output feature map. If there are multiple input channels and convolutional kernels, each kernel executes convolution with its associated channel. Then, the results are summed to form the final output feature map as shown in Fig. \ref{fig:traditionalDepthSeparableCompare}(a). According to such a routine procedure to perform convolution, the computation load would be extremely expense. 

To address this issue, as shown in Fig. \ref{fig:traditionalDepthSeparableCompare}(b), a depthwise-separable convolution fashion is figured out in consideration of FHE, that is, the traditional convolution is decomposed into two smaller parts: depthwise convolution and pointwise convolution, by which both the parameter amount and computation load are remarkably cut down. Depthwise convolution acts as the first stage of our proposed depthwise-separable convolution, wherein each input channel is convolved independently with a dedicated spatial kernel. Unlike traditional convolution, it employs the same number of kernels as input channels, with each kernel operating solely on its associated channel and no cross-channel fusion occurring. This design significantly reduces computation load, as the convolution of each channel is processed independently while preserving the original channel count. Pointwise convolution constitutes the second stage, utilizing $1$$\times$$1$ kernel to perform cross-channel feature fusion. Since the kernel size is $1$$\times$$1$, it retains the spatial dimension of the feature map while only modifying the channel depth, whose primary role is to linearly combine the multi-channel outputs from depthwise convolution, projecting them into a new channel space to enable efficient inter-channel communication with minimal computation load. 


\textbf{Analysis on Parameter Amount and Computation Load.} We formally analyze the saving amounts of parameters and computation load. Suppose the kernel size is $f$, the numbers of input and output channels are $c_i$ and $c_o$, the output feature map has a size of $h_o$$\times$$w_o$. For the traditional convolution, the number of parameters is $f$$\times$$f$$\times$$c_i$$\times$$c_o$, and the total computation amount is $f$$\times$$f$$\times$$c_i$$\times$$c_o$$\times$$h_o$$\times$$w_o$. However, for our proposed depthwise-separable convolution, the total number of parameters is $f$$\times$$f$$\times$$c_i+c_i$$\times$$c_o$, and the total computation amount becomes $f$$\times$$f$$\times$$c_i$$\times$$h_o$$\times$$w_o+c_i$$\times$$c_o$$\times$$h_o$$\times$$w_o$. Accordingly, the ratio of parameters between the proposed depthwise-separable convolution and the traditional convolution is:
\begin{equation}
	\frac{f\times f\times c_{i}+ c_{i}\times  c_{o}}{f\times f\times c_{i}\times c_{o}} = \frac{1}{c_{o}}+\frac{1}{f^{2} },
\end{equation}
and the ratio of computation load between them is:
\begin{equation}
	\frac{f\times f\times c_{i}\times h_{o}\times w_{o} + c_{i}\times c_{o}\times h_{o}\times w_{o}}{f\times f\times c_{i}\times c_{o}\times h_{o}\times w_{o}}=\frac{1}{c_{o}} +\frac{1}{f^{2} }. 
\end{equation}

When the output channel 
\begin{math}
	c_{o} 
\end{math}
is relatively large (e.g. 32 or 64), the term 
\begin{math}
	\frac{1}{c_{o} }  
\end{math}
can be ignored. If
\begin{math}
	f=3 
\end{math}
(i.e. a 
\begin{math}
	3 \times 3
\end{math}
convolution kernel), under our depthwise-separable convolution, the computational load drops to around $ \frac{1}{9}$  of that of the traditional convolution. Next, we will analyze how to apply in an encrypted environment.

\begin{figure*} [hbtp]
	\centering
	\subfigure[Depthwise convolution]{
		\begin{minipage}[t]{0.95\textwidth}
			\centering
			\includegraphics[width=6.5in, height=0.9in]{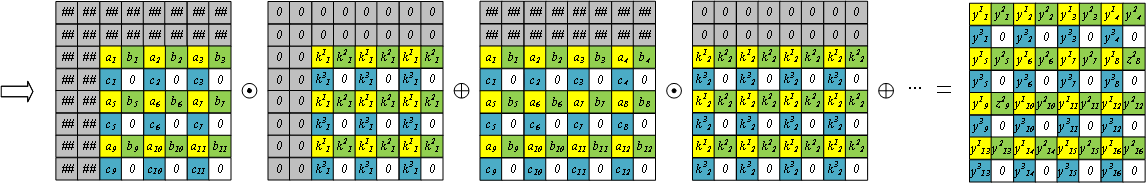}
			\label{Fig:sub1}
		\end{minipage}
	}
	\vspace{0.02cm}
	\subfigure[Pointwise convolution]{
		\begin{minipage}[t]{0.95\textwidth}
			\centering
			\includegraphics[width=3.8in, height=0.9in]{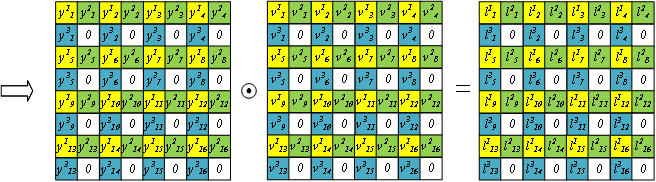}
			\label{Fig:sub2} 
		\end{minipage}
	}
	\vspace{0.01cm}
	\subfigure[Accumulation of the input-channel convolution results to form a single output-channel feature map]{
		\begin{minipage}[t]{0.95\textwidth}
			\centering
			\includegraphics[width=6.5in, height=0.9in]{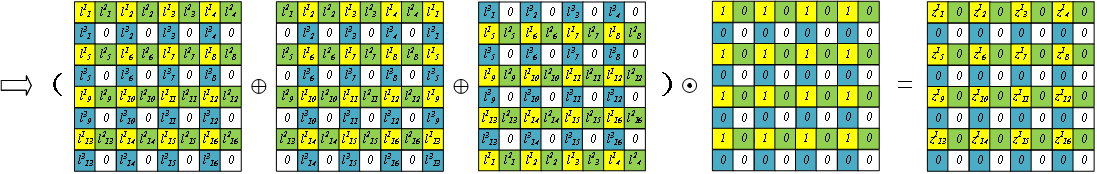}
			\label{Fig: PGD} 
		\end{minipage}
	}
	\vspace{0.01cm}
	\subfigure[Aggregation of multiple output-channel convolution results]{
		\begin{minipage}[t]{0.95\textwidth}
			\centering
			\includegraphics[width=5.2in, height=0.9in]{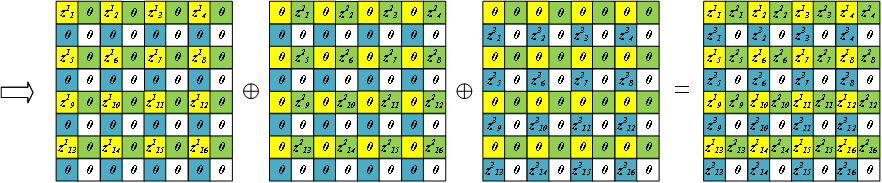}
			\label{Fig: PGD} 
		\end{minipage}
	}
	\caption{Block-ciphertext depthwise-separable multi-channel convolution.}
	\label{fig:depthSepaConvo}
\end{figure*}

\textbf{Quantifying Ciphertext Computation in Traditional Convolution.} In the ciphertext domain, the traditional convolution process is illustrated in Fig. \ref{fig:ciphertextTraditionalConvo}  in Appendix \ref{Sec:traditionCiphertextConvo}. At first, both the encrypted input data and the encoded kernel plaintext must be rotated, which requires $(f^2-1)$ rotations. Then, the corresponding ciphertext and plaintext are multiplied and summed to compute the convolution results across multiple input channels, which involves $f^2$ multiplications and consuming $1$-level multiplicative depth. Afterward, these multi-channel convolution results are summed to generate the output feature map for a single output channel, which requires $c_i$ rotations and $1$ multiplication, and consumes one more multiplicative depth. Thus, for one output channel, the total rotation cost is $f^2-1+c_i$, the total multiplication cost is $f^2+1$, and the total levels of multiplicative depth cost is $2$. To produce $c_o$ output channels, the required operations become $(f^2-1+c_i )$$\times$$c_o$ rotations, $(f^2+1)$$\times$$c_o$ multiplications and $2$-level multiplicative depth.
Fig. \ref{fig:depthSepaConvo} sketches the procedure of block-ciphertext depthwise-separable multi-channel convolution, and the more concrete description is illuminated in Appendix \ref{Sec:depthSepaCiphertextConvo}.

\textbf{Quantifying Ciphertext Computation in Depthwise-Separable Convolution.}
In our depthwise-separable convolution under ciphertext scenario, the encrypted input data and encoded kernel plaintext at first undergo a depthwise convolution, which requires $(f^2-1)$ rotations and $f^2$ multiplications. Afterward, the pointwise convolution is performed, with the cost of $c_i$ rotations and $2$ multiplications. For $c_o$ output channels, the process in total uses ($f^2-1+c_i$$\times$$c_o$) rotations and ($f^2+2c_o$) multiplications. Therefore, in the ciphertext environment, the ratio of required rotations between our proposed depthwise-separable convolution and the traditional convolution is:
\begin{equation}
	\frac{\text{Depthwise-Separable Convolution}}{\text{Traditional Convolution}}
	= \frac{f^{2}-1+c_{i}\times c_{o}   }{(f^{2}-1+c_{i} )\times c_{o}}, 
\end{equation}
and the ratio of multiplications between them is: 
\begin{equation}
\label{Equ-MultiRatio}
\frac{f^{2} +2c_{o} }{(f^{2} +1)\times c_{o}} =\frac{f^{2} }{(f^{2} +1)\times c_{o}} +\frac{2}{f^{2}+1 }\approx \frac{2}{f^{2}+1 }.
\end{equation}

It is clear that as the number of output channels increases, the depthwise-separable convolution achieves an increasingly pronounced reduction on computation load. Given that the beginning conventional layers capture the shallow features of the input samples, such as edges, textures, and colors, which play important roles for the downstream tasks, hence, we apply such block-ciphertext depthwise-separable multi-channel convolution fashion to those layers that have a large number of channels. When $c_o$ is large, the $\frac{f^{2} }{(f^{2}+1 )\times c_{o} }$ term can be neglected, the number of multiplications is cut down to $\frac{1}{5} $ ($f=3$) of the original approximately. Furthermore, regrading the parameters of kernel plaintext, for instance, a $32$$\times$$16$$\times$$16$ tensor would require $f^2$$\times$$c_o$$=9$$\times$$32=288$ plaintexts of convolution kernel (parameter files) in the traditional convolution, but only $f^2+c_i$$=9+32=41$ plaintexts in our proposed depthwise-separable convolution.

\subsection{Fusing Convolutional Layer with BN Layer}
As shown in Fig. \ref{fig:depthSepaConvo}, splitting the single-step convolution process into two parts of (depthwise and pointwise) convolutions entails one additional multiplicative depth. In FHE, the utilization of multiplicative depth must be carefully planned to control both the placement and frequency of bootstrapping with the purpose of keeping the added overhead within a reasonable range. Taking into account that the convolutional layer and batch-normalization (BN) layer are both the combination of multiplication and addition, one question appears naturally, whether we can merge them together to reduce multiplications and multiplicative depths or not. To this end, we propose a BN dot-product fusion matrix with the purpose of fully taking advantage of the homomorphic multiplication and associated multiplicative depth in the convolution layers. That is to say, we can merge the convolutional layer and BN layer while simultaneously eliminating the extra multiplicative depth incurred by the proposed depthwise-separable convolution operations.

To accomplish the fusion, we first reshape the BN layer's calculation formula into a form that consists of multiplication and addition operations only, yielding:
\begin{align}
y_i &= \gamma \hat{x}_i + \beta 
= \gamma \frac{x_i - \mu_B}{\sqrt{\sigma_B^2 + \epsilon}} + \beta \notag \\ 
&= \frac{\gamma}{\sqrt{\sigma_B^2 + \epsilon}} (x_i - \mu_B) + \beta 
= \alpha \mathbb{X}_{i}  + \beta \label{eq:middle},
\end{align}
where 
\begin{math}
	\alpha =\frac{\gamma }{\sqrt{\sigma _{B}^{2}+\epsilon  } } 
\end{math} 
represents the BN layer's coefficient, and 
\begin{math}
	\mathbb{X}_{i} =x_{i} -\mu _{B} 
\end{math} 
is the result after subtracting the batch mean from the convolution output. In this way, we obtain a BN layer formula that is expressed solely in terms of addition and multiplication. Specifically, 
\begin{math}
	x_{i}
\end{math} 
is first subtracted by the mean 
\begin{math}
	\mu _{B}
\end{math}, 
then multiplied by the coefficient 
\begin{math}
	\alpha
\end{math}, 
and finally the bias 
\begin{math}
	\beta 
\end{math}
is added.

\textbf{ConvBN without Multiplicative-Depth Optimization.} One necessary manipulation is to integrate the BN layer's computation into the convolutional layer's calculation process. As demonstrated in Fig. \ref{fig:depthSepaConvo}, after the execution of pointwise convolution, the multi-input channel convolution results are accumulated to form a single output-channel result, and then a mask is applied to filter out the redundant data produced by ciphertext rotations. This process consumes two levels of multiplicative depth. Our approach aims to merge these steps into one and incorporate the BN layer, forming a ConvBN layer. Appendix \ref{Sec:ConvBNWithoutOptimization} illustrates the ConvBN process without the multiplicative-depth optimization. Compared with Fig. \ref{fig:depthSepaConvo}, the only modification lies in that the result obtained from the pointwise convolution is first subtracted by the mean matrix 
\begin{math}
	\mu _{B}
\end{math},
then multiplied by the BN coefficient $\alpha$, which combines the mask matrix with the BN layer's coefficient matrix, finally, during the aggregation of multi-output channel results, the offset matrix 
\begin{math}
	\beta
\end{math}
is added.
%

\begin{figure*} [htbp]
	\centering
	\subfigure[BN dot-product fusion matrix]{
		\begin{minipage}[t]{0.2\textwidth}
			\centering
			\includegraphics[width=1.0in, height=0.8in]{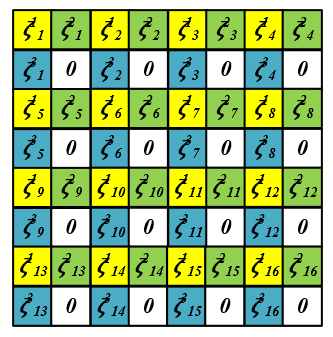}
			\label{Fig:sub1}
		\end{minipage}
	}
	\vspace{0.01cm}
	\subfigure[Generate rotation matrices concordant to input channels and filter out redundant data]{
		\begin{minipage}[b]{0.6\textwidth}
			\centering
			\includegraphics[width=3.3in, height=0.9in]{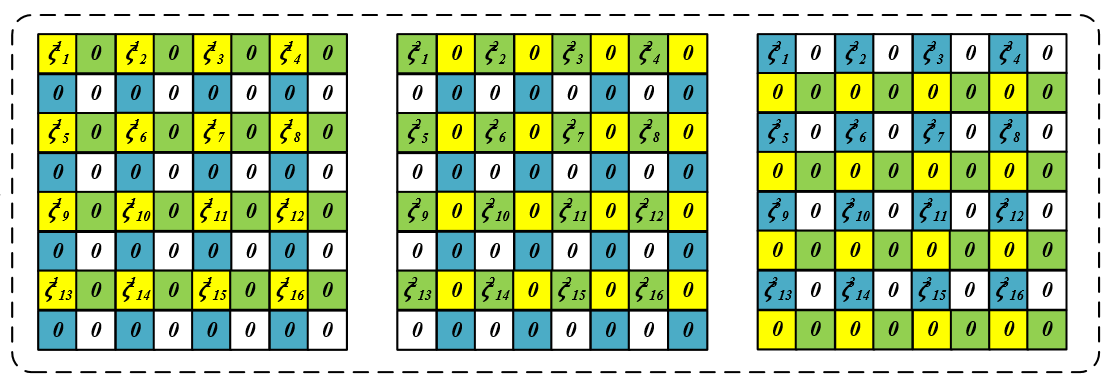}
			\label{Fig:sub2} 
		\end{minipage}
	}
	\vspace{0.01cm}
	\subfigure[Fusion of steps (b) and (c) in Fig. \ref{fig:depthSepaConvo} with optimized multiplicative depth consumption]{
		\begin{minipage}[b]{0.95\textwidth}
			\centering
			\includegraphics[width=5.4in, height=2.5in]{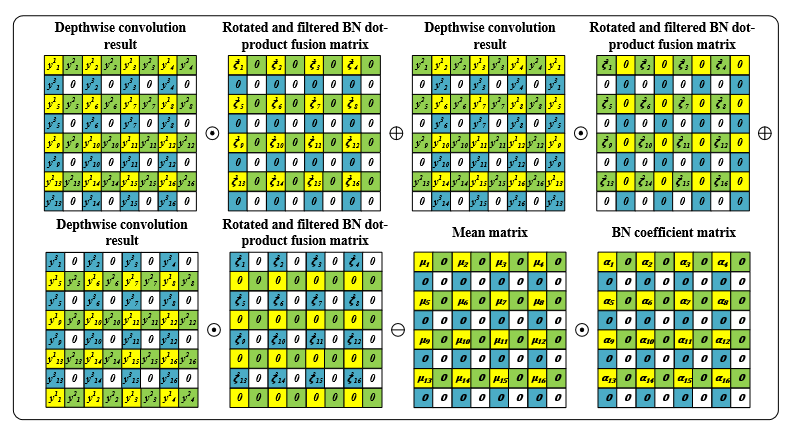}
			\label{Fig: PGD} 
		\end{minipage}
	}
	\caption{BN dot-product fusion matrix and ConvBN multiplicative depth optimization.}
	\label{fig:BNDotProductFusion}
\end{figure*}

\textbf{ConvBN with Multiplicative-Depth Optimization.} On the basis of Fig. \ref{fig:ConvBNWithNoOptimizaiton} in Appendix \ref{Sec:ConvBNWithoutOptimization}, we can further optimize the multiplicative depth so that only one level is consumed. In detail, our approach is to fuse the parameters of pointwise convolution with the BN coefficient $\alpha$ to form a new coefficient matrix—the BN dot-product fusion matrix. Then, through performing the ciphertext rotations, multiplications and accumulations between the depthwise convolution and BN dot-product fusion matrix, we can simultaneously accumulate the convolution results from different input channels and apply BN manipulation for the regularization purpose, the procedure is described in Fig. \ref{fig:BNDotProductFusion}. 
Therefore, our proposed BN dot-product fusion matrix achieves the aggregation of pointwise convolution, BN coefficient $\alpha$ and mask to filter out redundant data.


Consequently, the element of BN dot-product fusion matrix is the product of the pointwise convolution kernel parameter and the BN coefficient, namely, 
\begin{equation}
	\zeta _{n}^{i} = \nu _{n}^{i} \times \alpha _{o} =\nu _{n}^{i}\times \frac{\gamma }{\sqrt{\sigma _{B}^{2}+\epsilon  } }, 
\end{equation}
where
\begin{math}
	\alpha _{o}
\end{math}
denotes the BN coefficient for each output channel's BN layer, and 
\begin{math}
	\nu _{n}^{i}
\end{math}
represents the pointwise convolution kernel parameter for the 
\begin{math}
	i
\end{math}th input channel at position 
\begin{math}
	n
\end{math}.
After generating the BN dot-product fusion matrix, we need to perform both rotation and filtering on this matrix. The rotation, performed according to the number of input channels, serves to shift the parameters of each channel to the first position within the 
\begin{math}
	N \times N
\end{math}
block. Consequently, when the depthwise convolution result undergoes the corresponding rotation, multiplying it by the rotated fusion matrix directly produces the result of the pointwise dot-product, followed by the BN coefficient multiplication for that channel, and the result is placed in the first position within the block to facilitate the accumulation of multi-input channel convolution results. Furthermore, the filtering step is designed to eliminate any redundant data brought by the rotations of depthwise convolution. 
We provide the schematic diagrams of our proposed ConvBN with/without optimization on multiplicative depth consumption  in Appendix \ref{Sec:ConvBNDiagram}. 

Since the parameters of BN dot-product fusion matrix are derived from plaintext, this matrix does not incur any limitation in terms of multiplicative depth consumption or rotation count, that is to say, it can be generated directly in the plaintext domain. The actual consumption of multiplicative depth and rotations within the ciphertext is confined to the content inside the solid-line box in Fig. \ref{fig:BNDotProductFusion}. This portion consolidates the two steps in Fig. \ref{fig:depthSepaConvo}(b)-(c) of the block-ciphertext depthwise-separable convolution into a single step by utilizing rotated and filtered BN dot-product fusion matrix. As a result, under the same rotation count, one multiplicative depth operation can be saved. 

\section{Activation Function Approximation}
Since FHE only handles the homomorphic multiplication and addition operations, scholars usually leverage polynomial to approximate the nonlinear activation functions. At present, the pioneering CryptoDL \cite{hesamifard2018privacy} and other subsequent works \cite{ishiyama2020highly, lou2021hemet, obla2020effective, hesamifard2019deep} have employed different-degree polynomials to approximate the activation functions, however, few of them have demonstrated the effectiveness (inference accuracy) and rationality in consideration of both efficiency and approximation precision. Upon the in-depth investigation on existing studies, we find the challenge lies in the following predicament: the low-degree polynomial cannot precisely approximate the activation function, further leading to a significant drop in model inference accuracy; while the high-degree polynomial consumes a huge amout of multiplicative depth, enforcing frequent bootstrapping operations in residual blocks in spite that a high inference accuracy can be gained. Therefore, an ideal solution is to seek a lightweight solution that enables to precisely approximate the nonlinear activation function without consuming too much multiplicative depth at the same time.

At present, the high-degree polynomial typically employs Chebyshev polynomial to approximate ReLU function, while the low-degree polynomial usually utilizes Hermite polynomial. As well known, the polynomial-approximation technique can perform well on smooth functions, whereas the ReLU function has a discontinuity point, it is not a smooth function. To gain an accurate approximation, the Chebyshev polynomial must be high-degree. Oppositely, the Hermite polynomial with low-degree inevitably sacrifices inference accuracy. To address the dilemma, we replace ReLU with SiLU, a continuous function over real number domain 
\begin{math}
	\mathbb{R}
\end{math} and mathematically expressed as, 
\begin{equation}
	SiLU(x)=x\cdot \sigma (x)=\frac{x}{1+e^{-x} }, 
\end{equation}
where 
\begin{math}
	\sigma (x)=\frac{1}{1+e^{-x} } 
\end{math} 
denotes the standard Sigmoid function. Since it provides gradient over its entire range, especially in the negative region, it prevents the neurons from completely ceasing to update. Consequently, it tends to perform better than ReLU and its variants (Leaky ReLU).

Aiming to realize lightweight computation, we adopt low-degree Legendre polynomials to approximate SiLU function, stemming from the property of smoothness. The goal of Legendre polynomial approximation is to minimize the mean squared error, that is, to minimize the integral of the squared error over the entire range:
\begin{equation}
	err=\int_{-1}^{1} (f(x)-\sum_{k=0}^{n}a_{k}P_{k}(x))^{2} dx,
\end{equation}
where $f(x)$ denotes activation function, and $\sum_{k=0}^{n}a_{k}P_{k}(x)$ indicates the polynomial for approximation. 
Thus, it provides a better average approximation over the entire interval, and this excellent performance across a wide range is clearly more suitable for representing nonlinear activation functions.

Legendre polynomials are a class of orthogonal polynomials that play a crucial role in mathematics and physics, typically denoted as 
\begin{math}
	P_{n}(x)
\end{math}, 
where 
\begin{math}
	n
\end{math} 
represents the degree of the polynomial. These polynomials can be computed using the following recurrence relation:
\begin{equation}
	P_{n}(x)=\left\{\begin{matrix}
		1&n=0 \\
		x&n=1 \\
		\frac{(2n+1)xP_{n}(x)-nP_{n-1}(x)}{n+1} &n\ge 2
	\end{matrix}\right ..
\end{equation}

For a given function 
\begin{math}
	f(x)
\end{math}, 
we can express it as a linear combination of Legendre polynomials:
\begin{equation}
	f(x)=\sum_{n=0}^{\infty}a_{n}P_{n}(x),   
\end{equation}
where
\begin{math}
	a_{n}
\end{math} 
is determined by the following formula:
\begin{equation}
	a_{n} =\frac{2n+1}{2}\int_{-1}^{1}f(x)P_{n}(x)dx.   
\end{equation}

\begin{figure}[bthp] 
	\centering
	\includegraphics[width=2.0in, height=1.4in]{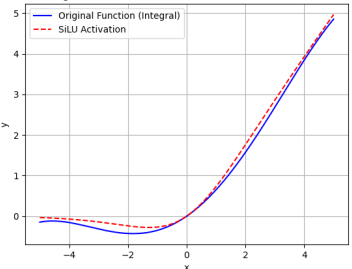}
	\caption{Polynomial approximation of SiLU}
	\label{fig:SiLUApproximation}
\end{figure}

By computing sufficiently many coefficients 
\begin{math}
	a_{n}
\end{math}, we can obtain an approximation polynomial with a finite number of terms. Since our goal is to achieve a precise approximation and consume minimal multiplicative depth, we used the formula above to derive a polynomial of maximum degree 5 to approximate SiLU. This approximation consumes 3 levels of multiplicative depth, as shown in Fig. \ref{fig:SiLUApproximation}.

\section{Reshaping Network Architecture over FHE}
\begin{figure*}[hbtp]
	\centering
	\includegraphics[width=6.6in, height=2.2in]{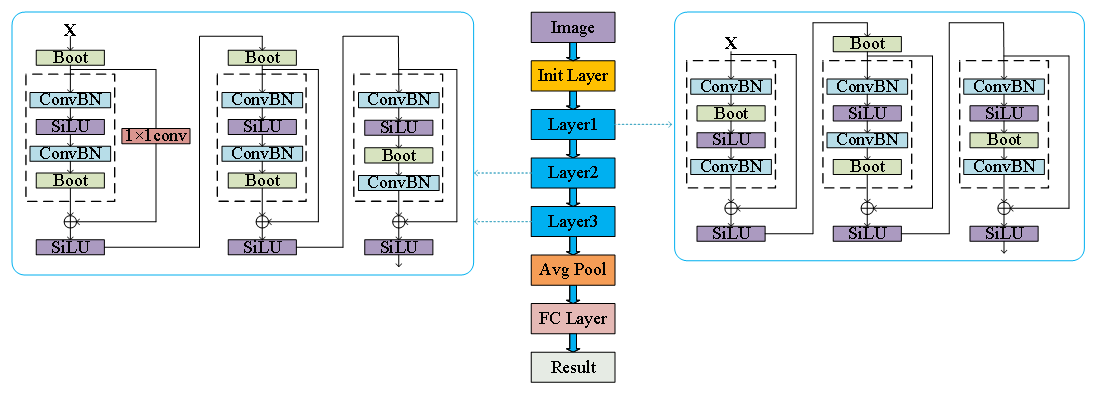}
	\caption{Rebuilt ResNet20 architecture in consideration of FHE.}
	\label{fig:ResNetArchiFHE}
\end{figure*}

By integrating our proposed ConvBN and employed activation function SiLU, and subsequently incorporating downsampling, AVGPOOL, fully connected layers, we need reshape the network architecture using RNS-CKKS scheme, and take ResNet20 as an example as shown in Fig. \ref{fig:ResNetArchiFHE}.

Since the traditional convolution can fully extract the shallow features and our ConvBn is especially effective to reduce computation load when the number of channels is large, we use conventional convolution in Layer 1 and adopt the proposed ConvBn in Layer 2 and Layer 3. We determine the placement of bootstrapping based on two factors: i) whether the remaining multiplicative depth is insufficient to support the following operation, otherwise the bootstrapping must be performed; and ii) ensuring the ciphertexts involved in skip connections are at the same multiplicative depth as much as possible, otherwise the addition of ciphertexts at different multiplicative depths would incur computation error or decryption failure. 

\section{Performance Evaluation}
\subsection{Configuration}
To validate the efficiency and effectiveness of our proposed FastFHE, we perform multi-facet experiments on the commonly-used dataset CIFAR-10 \cite{krizhevsky2009learning} using RNS-CKKS scheme from OpenFHE Library \cite{al2022openfhe}. 
The detailed configurations on the running environment, dataset and models, settings on encryption-parameters, statistics of three baselines and our FastFHE, are demonstrated in Appendix \ref{Sec:Configuration}.  

\subsection{Resource Consumption Performance}
For the encrypted-data computation, the consumption of various resources is an important evaluation criterion, we assess the performance in terms of ciphertext size, runtime memory usage, and key size. As shown in Table \ref{tab:spaceResource}, the pioneering work \cite{lee2022privacy} leverages single-ciphertext and single-channel (SCSC) data-packing scheme to apply into PPML domain, while AESPA \cite{park2022aespa} and AutoFHE \cite{ao2024autofhe} use single-ciphertext and multi-channel (SCMC) data-packing manner. Easy to know, the SCSC data-packing scheme dominates more computing resources compared to the SCMC, this is because the ciphertext amount will be enlarged as the number of channels increases, for example, the former consumes 376MB to store ciphertext and 473 GB memory to run procedure, while the latter consumes much smaller storage spaces, i.e. 8.5 MB for AESPA and 7.2MB for AutoFHE, and uses smaller memories, i.e. 316 GB and 270GB respectively. Compared to these baselines, our FastFHE consumes the least resources, i.e. 6.3MB for ciphertext storage and 80GB memory for procedure running. Moreover, our approach significantly reduces the key size to less than $\frac{1}{2}$ of that of the best baseline.   

\begin{table*}[htbp]
	\centering
	\caption{Resource consumption with ResNet20.}
	\label{tab:spaceResource}
	\begin{tabular}{ccccc}  
		\toprule
		\textbf{Model} & \textbf{Packing Method} & \textbf{Ciphertext Size} & \textbf{Memory Usage} & \textbf{Key Size} \\
		\midrule
		Literature \cite{lee2022privacy} & Single-ciphertext, single-channel & 376MB & 472GB & 109GB \\
		AESPA\cite{park2022aespa} & Single-ciphertext, multi-channel & 8.5MB & 316GB & 133GB\\
		AutoFHE\cite{ao2024autofhe} & Single-ciphertext, multi-channel & 7.2MB & 270GB & 121GB\\
		FastFHE & \begin{math}
			N \times N 
		\end{math} 
		Block based packing & \textbf{6.3}MB & \textbf{80}GB & \textbf{42.1}GB \\
		\bottomrule
	\end{tabular}
\end{table*}

Furthermore, AESPA\cite{park2022aespa} and AutoFHE\cite{ao2024autofhe} mainly focus the attention on the approximation of activation layers, ignoring the pivotal correlation between data arrangement and rotation manipulation. This oversight would induce excessively high memory consumption, deriving from the huge number of ciphertext rotations to calculate convolution in the pipeline of neural network, in addition to the loading of rotation key and the homomorphic multiplication operations. To address this issue, our work proposes the scalable ciphertext data-packing scheme, it enables the multi-channel encrypted data to be arranged within each block in a particular rule, by which both the number of ciphertext rotations and the amount of rotation keys can be remarkably reduced.

\begin{figure}[hbtp] 
	\centering
	\includegraphics[width=2.6in, height=1.4in]{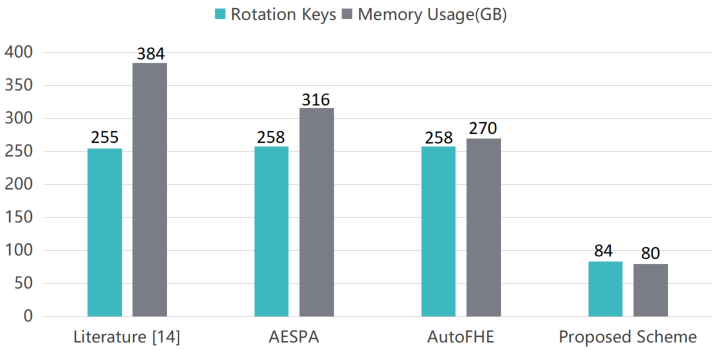}
	\caption{Rotation keys and memory usage.}
	\label{fig:RotationMemoryCount}
\end{figure}

In addition, our proposed data-packing scheme can also achieve the simultaneous shifting of multiple channels in a single ciphertext rotation, greatly declining the memory overhead. As shown in Fig. \ref{fig:RotationMemoryCount}, compared to the baselines, our FastFHE needs the smallest amount of rotation keys and memory usage under ResNet20. On the other hand, during the ciphertext convolution operations, the rotated ciphertexts produced by homomorphic rotations also occupy additional memory space, the more rotations performed, the larger the memory footprint of these rotated ciphertexts. These two factors are the fundamental causes of high memory consumption observed in the three baseline models.

\subsection{Overhead in Depthwise-Separable Convolution}
This section compares our proposed block-ciphertext depthwise-separable convolution (BCDSC) with the traditional ciphertext convolution. In the early stages of inference using ResNet20, the input features are abstract, and the neural network needs to fully capture both spatial and inter-channel characteristics to uncover correlations between different channels. Traditional convolution is more effective at extracting inter-channel information. Therefore, ciphertext convolution is employed at the initial layer and at Layer 1. In the later stages, where the number of channels increases significantly, the proposed block-ciphertext depthwise-separable convolution is adopted to reduce computational complexity and parameter amount. This treatment aims to improve convolution efficiency in ciphertext domain while maintaining inference accuracy at the same time.

Under the ciphertext environment, we process CIFAR-10 test data using our scalable $N$$\times$$N$ block-oriented data-packing method. The proposed BCDSC is applied at Layer 2 and Layer 3, and the activation functions are approximated by a degree-6 Chebyshev polynomial resembling the ReLU function, ensuring compatibility with ciphertext-domain CNN models. Under both traditional convolution and our BCDSC, Table \ref{tab:ConvoCalTime} and Table \ref{tab:plaintextPara} detail the statistics for each layer and total time for convolution operations, as well as the number of plaintext convolution kernel parameters involved. Since the number of parameters directly correlates with the number of multiplications, it also indirectly reflects the computation advantage of our proposed method.

\begin{table*}[htbp]
	\centering
	\caption{Convolution calculation time.}
	\label{tab:ConvoCalTime}
	\begin{tabular}{cccccc}  
		\toprule
		\textbf{Convolution Algorithm} & \textbf{Init. Layer} & \textbf{Layer 1} & \textbf{Layer 2} & \textbf{Layer3 } & \textbf{Total} \\
		\midrule
		Traditional convolution method & 8s & 54s & 472s & 754s & 1,327s \\
		Block ciphertext depthwise separable convolution method & 8s & 54s & 97s & 150s & 309s\\
		\bottomrule
	\end{tabular}
\end{table*}

\begin{table*}[htbp]
	\centering
	\begin{threeparttable}
		\caption{Number of plaintext convolution kernel parameters.}
		\label{tab:plaintextPara}
		\begin{tabular}{cccccc}  
			\toprule
			\textbf{Convolution Algorithm} & \textbf{Init. Layer} & \textbf{Layer 1} & \textbf{Layer 2} & \textbf{Layer 3} & \textbf{Total} \\
			\midrule
			Traditional convolution method & 144 & 864 & 1782+32 & 3,456+64 & 6,342 \\
			Block ciphertext depthwise separable convolution method & 144 & 864 & 246+32 & 438+64 & 1,788\\
			\bottomrule
		\end{tabular}
		\begin{tablenotes}
			\footnotesize
			\item Note: "+32" and "+64" at Layer 2 and Layer 3 denote the plaintexts for downsampling convolutions in skip connections
		\end{tablenotes}
	\end{threeparttable}
\end{table*}

During traditional convolution, the multi-channel ciphertexts packed in $N$$\times$$N$ blocks need to generate $f^2$ rotated ciphertexts. These rotated ciphertexts are multiplied and accumulated with the corresponding $N$$\times$$N$ block packed multi-channel plaintext kernels to compute the convolution results across input and output channels. Therefore, for one convolutional layer, the traditional method requires $f^2$$\times$$c_o$ plaintexts of convolution kernels. Specifically, Layer 2 requires 1782 plaintexts, and Layer 3 requires 3456 plaintexts. In contrast, under the proposed BCDSC scheme, deep convolution only requires $f^2$ plaintexts, while pointwise convolution requires $c_o$ plaintexts. The total plaintext in Layer 2 is $(f^2+c_o)$$\times$$n_{conv}=246$, and in Layer 3 is 438. As shown in Table \ref{tab:plaintextPara}, since the initial layer and Layer 1 still use traditional convolution under ciphertext, the number of plaintexts required in those stages is the same. However, in Layer 2 and Layer 3, the proposed method significantly reduces the number of plaintexts, which also implies a substantial reduction in the number of homomorphic multiplication operations.

As previously analyzed in Equ. \ref{Equ-MultiRatio}, the number of multiplications required by our scheme is approximately $\frac{2}{f^{2} +1} = \frac{1}{5} $ of that by the traditional convolution. In Table \ref{tab:plaintextPara}, the plaintext amount under our scheme in Layer 2 and Layer 3 is reduced by approximately 5 times in theory compared to traditional convolution (by 6.52$\times$ and 7.01$\times$ experimentally, due to the different number of external multiplications required for matrix-vector multiplication in the calculation). Furthermore, as shown in Table \ref{tab:ConvoCalTime}, the execution time for our proposed BCDSC in Layer 2 and Layer 3 is only $\frac{1}{4.86}$ and $\frac{1}{5.03}$ of that of traditional ciphertext convolution. This fully demonstrates the advantages of our proposed block-ciphertext depthwise-separable multi-channel convolution fashion in terms of inference efficiency.

\subsection{Model Inference Latency}
Upon the optimization on ConvBN multiplicative depth and low-degree Legendre approximation function, we employ ResNet20, ResNet32, ResNet44, and VGG11 \cite{KarenSimonyan2015} to execute our proposed FastFHE. Table \ref{tab:amortizeRuntime} presents the inference latency and amortized runtime, wherein the fastest variant of AutoFHE \cite{ao2024autofhe} is used in our experiments. 

Seen from the results, we have the observations: i) it is evident MPCNN exhibits longer inference latency and amortized runtime compared to AESPA and AutoFHE, this is primarily because MPCNN does not pay specific attention on activation function approximation. It employs a high-degree (6, 7 or 8) Chebyshev polynomial function to approximate ReLU function; ii) in contrast, AESPA employs the low-degree (2) Hermite polynomial to approximate ReLU function. AutoFHE leverages layered mixed-degree polynomial approximations for the substitution of activation functions. It is the low/mixed-degree polynomial that decreases the multiplicative depth, then it reduces the frequency of bootstrapping operations. AutoFHE has similar inference latency and amortized runtime with AESPA, however, whose accuracy loss is less than AESPA as shown in Table \ref{tab:AccuracyError}, stemming from the variable polynomial's degree settings; and iii) although AESPA and AutoFHE focus on lightweight treatment on the activation layers, they ignore the long-time inference problem in linear layers. Differently, through adopting the low-degree Legendre polynomials to approximate the employed SiLU in our work, FastFHE incurs lower time consumption in the activation layer, furtherly, coupled with the efficient convolution based on scalable data-packing scheme, our FastFHE significantly outperforms the baselines in both inference latency and amortized runtime. Specifically, in a 30-thread environment, our amortized runtime is reduced by 3.19$\times$, 2.46$\times$, and 2.38$\times$ compared to MPCNN, AESPA, and AutoFHE, respectively.

\begin{table*}[t]
	\centering
	\caption{Amortized runtime per image under multiple threads (30 threads).}
	\label{tab:amortizeRuntime}
	\begin{tabular}{lccccccccc}
		\toprule
		\multicolumn{2}{c}{\textbf{Backbone}} & 
		\multicolumn{2}{c}{\textbf{MPCNN}} & 
		\multicolumn{2}{c}{\textbf{AESPA}} & 
		\multicolumn{2}{c}{\textbf{AutoFHE}} & 
		\multicolumn{2}{c}{\textbf{FastFHE}} \\
		\cmidrule(lr){1-2} \cmidrule(lr){3-4} \cmidrule(lr){5-6} \cmidrule(lr){7-8}\cmidrule(lr){9-10}
		\textbf{Model} & \textbf{Parameters} & \textbf{Latency} & \makecell{\textbf{Amortized}\\\textbf{Latency}} & \textbf{Latency} & \makecell{\textbf{Amortized}\\\textbf{Latency}} & \textbf{Latency} & \makecell{\textbf{Amortized}\\\textbf{Latency}} & \textbf{Latency} & \makecell{\textbf{Amortized}\\\textbf{Latency}} \\
		\midrule
		ResNet-20&269K&2487s&83s&1913s&64s&1842s&62s&\textbf{763}s&\textbf{26}s \\
		ResNet-32&464K&4438s&148s&3068s&101s&2944s&99s&\textbf{1463}s&\textbf{49}s \\
		ResNet-44&658K&6302s&210s&4775s&147s&4346s&145s&\textbf{2012}s&\textbf{75}s \\
		VGG-11&123K&1369s&46s&572s&19s&656s&22s&\textbf{272}s&\textbf{9}s \\
		\bottomrule
	\end{tabular}
\end{table*}

\textbf{Runtime of various Operation.} This subsection is to experimentally evaluate the execution time of various operations on the four reshaped neural network architectures. Table \ref{tab:varisouOperationRuntime} shows the time overhead for various operations, from which the following concludes can be obtained: i) for MPCNN, the activation function is approximated by high-degree polynomial, leading to much higher consumption compared to HerPN and EvoReLU. Moreover, to achieve high approximation precision, more frequent bootstrapping operations are required (accounting for 71\% to 74\% of the total inference time), which is the main reason for the high inference latency; ii) for AESPA, the time consumption for HerPN operations is lower. However, this low-degree polynomial results in remarkably reduced model inference accuracy as shown in Table \ref{tab:AccuracyError}, a trade-off unacceptable in exchange for shorter inference times, especially in the security-critical scenario (e.g. self-driving); iii) for the comparison purposes, we select the fastest variant of AutoFHE, where it exhibits linear layer latency similar to AESPA, but differs in the activation function, this is because AutoFHE employs a layered mixed-degree polynomial, and consumes slightly more time than AESPA, yet delivers a less accuracy loss compared to AESPA; iv) our FastFHE utilizes the depthwise-separable convolution, which drastically reduces the computation load of convolution operations, in addition to ConvBN optimization, as a result, its linear-layer time consumption is notably lower than that of other schemes. Additionally, by approximating SiLU function with low-degree Legendre polynomial, our FastFHE not only achieves an excellent approximation to maintain high inference accuracy, but also reduces the multiplicative depth consumption. This contributes significantly to lowering the time required for both activation functions and bootstrapping operations, thereby, further reducing the overall inference latency and amortized runtime.

\begin{table*}[htbp]
	\centering
	\caption{Runtime of various operations.}
	\label{tab:varisouOperationRuntime}
	\begin{tabular}{lcccccccccccc}
		\toprule
		\multirowcell{2}{\textbf{Backbone}} & 
		\multicolumn{3}{c}{\textbf{MPCNN}} & 
		\multicolumn{3}{c}{\textbf{AESPA}} & 
		\multicolumn{3}{c}{\textbf{AutoFHE}} & 
		\multicolumn{3}{c}{\textbf{FastFHE}} \\
		\cmidrule(lr){2-4} \cmidrule(lr){5-7} \cmidrule(lr){8-10}\cmidrule(lr){11-13}
		&\textbf{Linear} & \textbf{AppReLU} &\textbf{Boot} &\textbf{Linear} &\textbf{HerPN} &\textbf{Boot}&\textbf{Linear}&\textbf{EvoReLU}&\textbf{Boot}&\textbf{Linear}&\textbf{EvoSiLU}&\textbf{Boot} \\
		\midrule
		ResNet-20&417s&279s&1764s&899s&\textbf{23}s&870s&863s&49s&897s&\textbf{350}s&35s&\textbf{383}s \\
		ResNet-32&852s&448s&2929s&1539s&53s&1476s&1519s&91s&1333s&\textbf{647}s&\textbf{45}s&\textbf{747}s \\
		ResNet-44&906s&623s&4629s&2503s&\textbf{82}s&2490s&2472s&117s&1756s&\textbf{804}s&106s&\textbf{1102}s \\
		VGG-11&177s&167s&998s&347s&8s&226s&457s&6s&174s&\textbf{114}s&\textbf{5}s&\textbf{153}s \\
		\bottomrule
	\end{tabular}
\end{table*}

\begin{figure}[htbp] 
	\centering
	\includegraphics[width=2.8in, height=1.856in]{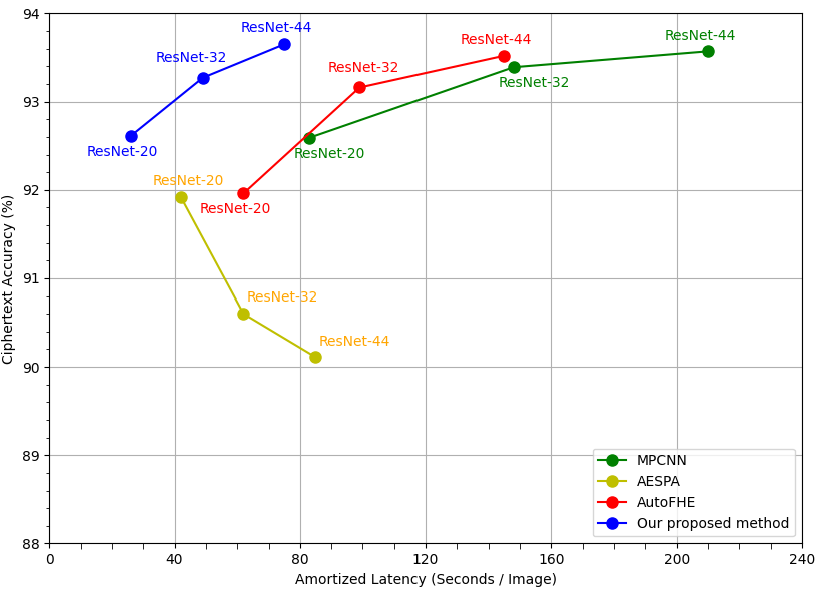}
	\caption{Cipertext accuracy and amortized latency.}
	\label{fig:accuracyLatency}
\end{figure}
\subsection{Model Inference Accuracy}
On the basis of comprehensive utilization of our proposed scalable data-packing scheme, ConvBN optimization, and low-degree Legendre polynomial approximation, we run a group of experiments to observe the overall comparison in terms of ciphertext accuracy and amortized latency on CIFAR-10 dataset under ResNet20/32/44. Obviously, our FastFHE located in the top-left corner in Fig. \ref{fig:accuracyLatency} has the best performance compared to the three baselines.  

\begin{table*}[htbp]
\centering
\caption{Accuracy and error between plaintext and ciphertext inference.}
\label{tab:AccuracyError}
\begin{tabular}{lcccccccccc}
	\toprule
	\multicolumn{2}{c}{\textbf{Backbone}} & 
	\multicolumn{2}{c}{\textbf{MPCNN}} & 
	\multicolumn{2}{c}{\textbf{AESPA}} & 
	\multicolumn{2}{c}{\textbf{AutoFHE}} & 
	\multicolumn{3}{c}{\textbf{FastFHE}} \\
	\cmidrule(lr){1-2} \cmidrule(lr){3-4} \cmidrule(lr){5-6} \cmidrule(lr){7-8}\cmidrule(lr){9-11}
	\textbf{Model} & \textbf{Accuracy} & \textbf{Accuracy} & \textbf{Error} & \textbf{Accuracy}& \textbf{Error}& \textbf{Accuracy}& \textbf{Error}& \textbf{Plaintext}& \textbf{Ciphertext}& \textbf{Error}\\
	\midrule
	ResNet20&92.76\%&92.59\%&0.17\%&91.92\%&0.84\%&91.96\%&0.80\%&92.66\%&92.61\%&\textbf{0.05\%} \\
	ResNet32&93.51\%&93.39\%&0.12\%&90.60\%&2.91\%&93.16\%&0.35\%&93.34\%&93.27\%&\textbf{0.07\%} \\
	ResNet44&93.92\%&93.57\%&0.35\%&90.11\%&3.81\%&93.52\%&0.40\%&93.86\%&93.65\%&\textbf{0.21\%} \\
	VGG11&90.53\%&90.41\%& \textbf{0.12\%}&88.97\%&1.56\%&90.69\%&0.16\%&90.22\%&90.01\%&0.21\% \\
	\bottomrule
\end{tabular}
\end{table*}

Although our proposed depthwise-separable convolutions in Layer 2 and Layer 3 can result in somewhat loss of inter-channel feature extraction, the efficient low-degree polynomial approximation of activation function, as well as the superior performance of smooth activation function SiLU under plaintext inference, both can compensate the accuracy. Moreover, due to the high-quality approximation of the activation function, the gap between plaintext accuracy and ciphertext accuracy is narrowed down to a tiny extent. As shown in Table \ref{tab:AccuracyError}, when using ResNet20 for inference, the accuracy error of our FastFHE before and after encryption is only 0.05, whereas the accuracy errors of MPCNN, AESPA, and AutoFHE are 0.17\%, 0.84\%, and 0.80\%, respectively. Similarly, under ResNet32 and ResNet44, our method also yields the smallest accuracy loss, i.e., 0.07\% and 0.21\%. For VGG11, the error is slightly higher than that of MPCNN and AutoFHE, this is primarily because ReLU outperforms our employed SiLU in the shallow layers.

\section{Related Work}
We review the existing work from two perspectives: i) the computation-load reduction of linear layer; and ii) the approximate expression of activation function of nonlinear layer. 
The detailed discussions and comparative analytics are presented in Appendix \ref{Sec:relatedWork}. 

\section{Conclusion}
We have experimentally and analytically validated the efficiency and effectiveness of our proposed FastFHE, a privacy-preserving deep neural network mechanism, through figuring out a scalable data-packing scheme, a lightweight yet highly-precise low-degree approximation polynomial, and a ConvBN architecture by fusing convolutional layer with BN layer. To conquer the grave drawbacks of tremendous bootstrapping and rotation operations, high resource consumption caused by the currently-prevalent stationary data-packing scheme, as well as the dilemma of time consumption and model inference accuracy in present works, we propose a novel scalable data-packing scheme to save the time consumption and storage resources, and employ depthwise-separable convolution fashion to reduce the computing load of convolution calculation. Furthermore, we design BN dot-product fusion matrix to integrate the convolution layer with the BN layer without incurring extra multiplicative depth. In addition, we identify a fundamental challenge in polynomial approximation on nonlinear activation functions lies in the difficulty of precisely approximating a non-smooth function like ReLU. Towards this issue, we adopted low-degree Legendre polynomial to approximately express the smooth function SiLU. Finally, extensive experiments demonstrate that, compared to the baselines MPCNN, AESPA and AutoFHE, our proposed FastFHE achieves the inference latency reductions of 3.26$\times$, 2.51$\times$, and 2.41$\times$ udner ResNet20, and in a 30-thread environment, the amortized runtime is reduced by 3.19$\times$, 2.46$\times$ and 2.38$\times$, respectively. 
From the additional experiments on ResNet32, ResNet44 and VGG11, the analogous analytics can be gained as well. Thus, our proposed lightweight PPML solution can boost the realistic applications in near future.

%

\bibliographystyle{IEEEtran}
\bibliography{sample-base}

\clearpage
\newpage
\appendices   

\section{Ciphertext Convolution under Traditional Operation} \label{Sec:traditionCiphertextConvo}
\begin{figure*} [hbtp]
	\centering
	\subfigure[Multiply and accumulate the rotated outputs]{
		\begin{minipage}[t]{0.95\textwidth}
			\centering
			\includegraphics[width=6.8in, height=2.2in]{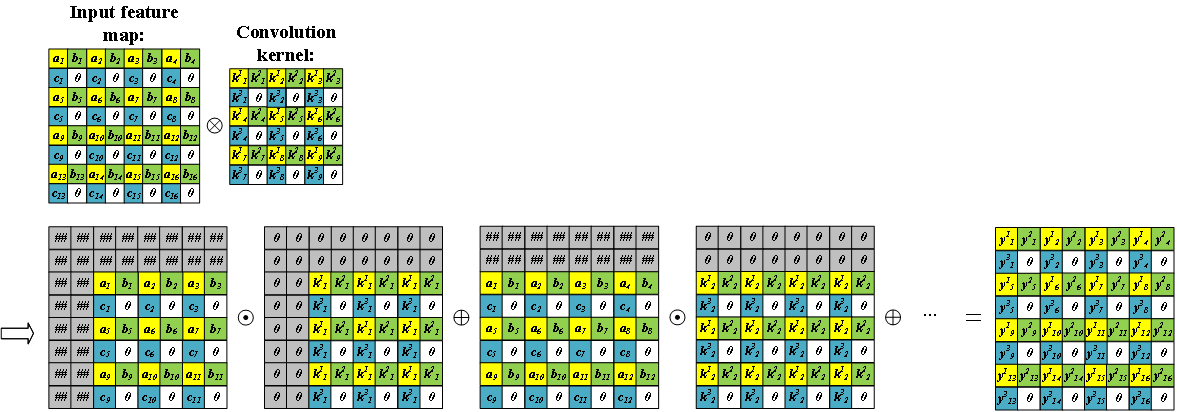}
			\label{Fig:sub1}
		\end{minipage}
	}
	\vspace{0.01cm}
	\subfigure[Accumulate results from the input channels to form a single output-channel feature map]{
		\begin{minipage}[b]{0.95\textwidth}
			\centering
			\includegraphics[width=6.8in, height=1.1in]{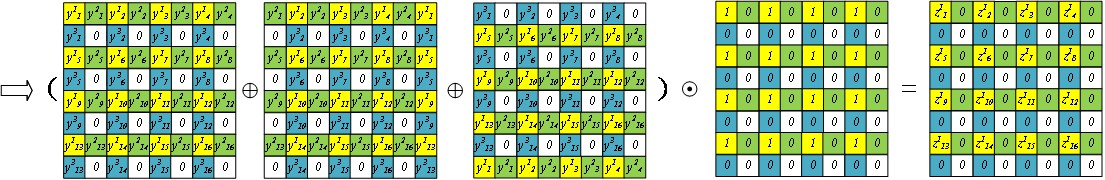}
			\label{Fig:sub2} 
		\end{minipage}
	}
	\vspace{0.01cm}
	\subfigure[Accumulate across all output channels to form the multi-channel convolution result]{
		\begin{minipage}[b]{0.95\textwidth}
			\centering
			\includegraphics[width=6.8in, height=1.1in]{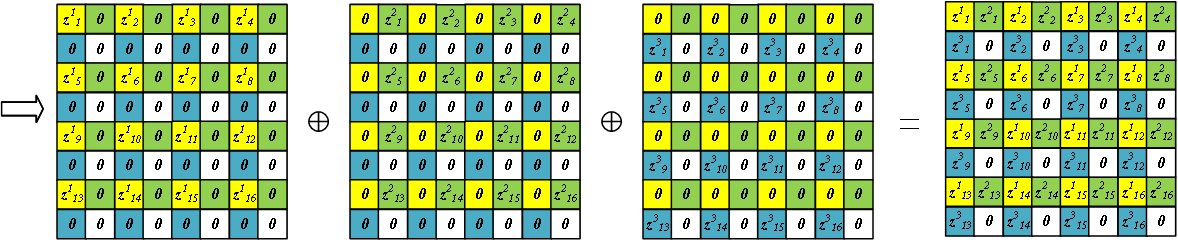}
			\label{Fig: PGD} 
		\end{minipage}
	}
	\caption{Multi-channel ciphertext convolution under traditional operations.}
	\label{fig:ciphertextTraditionalConvo}
\end{figure*}

Taking a 4$\times$4 feature map and 3$\times$3 convolution kernel as an example, as shown in Fig. \ref{fig:ciphertextTraditionalConvo}, the following demonstrates the detailed steps for the traditional convolution operation.

\begin{itemize}
\item 
\textbf{Step i):} Pack the feature map into pre-defined $N$$\times$$N$ blocks to form ciphertexts. Here, we use 2$\times$2 blocks for illustration. The multi-channel feature map is then transformed into the block units.

\item 
\textbf{Step ii):} Obtain the rotated ciphertexts for the kernel through rotating the ciphertext $f$$\times$$w$ times ($f$ and $w$ depend on the convolution kernel size). Multiply and accumulate to get the convolution results for the three input channels, as shown in Fig. \ref{fig:ciphertextTraditionalConvo}(a).	

\item 
\textbf{Step iii):} Combine the convolution results for the input channels to produce the output channel's feature map. The results are packed into $N$$\times$$N$ block positions concordant to the rotation offsets. A single reduction step is performed at this point to preserve only valid data through removing any extra rotation-induced artifacts, as illustrated in Fig. \ref{fig:ciphertextTraditionalConvo}(b).	

\item
\textbf{Step iv):} Repeat Steps i)-iii) for multi-channel settings. The combined results are placed into the associated block positions in ciphertext, generating an $N$$\times$$N$ ciphertext containing multi-channel output features, as shown in Fig. \ref{fig:ciphertextTraditionalConvo}(c).
\end{itemize}

\section{$\textbf{N}$$\times$$\textbf{N}$ Block Depthwise-Separable Multi-Channel Convolution} \label{Sec:depthSepaCiphertextConvo}
As illustrated in Fig. \ref{fig:depthSepaConvo}, the following presents the detailed procedure of depthwise-separable multi-channel convolution operations

\begin{itemize}
\item 
\textbf{Step i):} Multiply the ciphertext and kernel plaintext using the relevant rotation indices, then accumulate the results to obtain depthwise convolution outcome.
\item
\textbf{Step ii):} Perform pointwise convolution to produce the multi-input channel convolution result.
\item
\textbf{Step iii):} Accumulate the results of the multi-input channel convolution via rotations, generating the output feature map for a single output channel.
\item
\textbf{Step iv):} If multiple output channels are required, repeat Steps ii)-iii), placing the generated output into the concordant channel position in the $N$$\times$$N$ block. Ultimately, an $N$$\times$$N$ block-oriented ciphertext containing multiple output channels can be produced.
\end{itemize}

\section{ConvBN without Optimization} \label{Sec:ConvBNWithoutOptimization}
\begin{figure*} [htbp]
	\centering
	\subfigure[Pointwise Convolution]{
		\begin{minipage}[t]{0.95\textwidth}
			\centering
			\includegraphics[width=4.0in, height=1.1in]{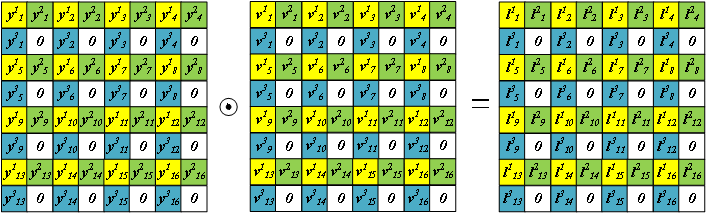}
			\label{Fig:TraditionalConv}
		\end{minipage}
	}
	\vspace{0.02cm}
	\subfigure[Accumulation of input-channel convolution results to form a single output-channel convolution result]{
		\begin{minipage}[t]{0.95\textwidth}
			\centering
			\includegraphics[width=5.8in, height=1.1in]{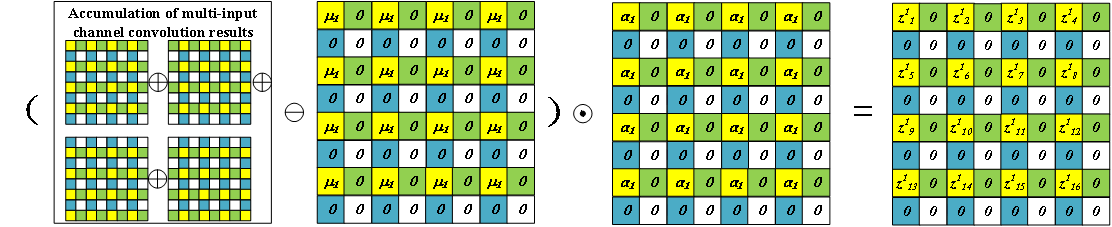}
			\label{Fig:DepthSepaConv} 
		\end{minipage}
	}
	\caption{ConvBN without optimization on multiplicative depth consumption.}
	\label{fig:ConvBNWithNoOptimizaiton}
\end{figure*}

As shows in Fig. \ref{fig:ConvBNWithNoOptimizaiton}, the ConvBN without optimization on multiplicative depth consumption is calculated as follows: the result obtained from the pointwise convolution is first subtracted by the mean matrix $\mu _{B}$, then multiplied by the BN coefficient matrix $\alpha$, which combines the mask matrix with the BN layer's coefficient matrix, finally, during the aggregation of multi-output channel results, the offset matrix $\beta$ is added.

\section{ConvBN Diagram} \label{Sec:ConvBNDiagram}
\begin{figure*} [htbp]
	\centering
	\subfigure[ConvBN without multiplicative-depth optimization]{
		\begin{minipage}[t]{0.45\textwidth}
			\centering
			\includegraphics[width=2.8in, height=2.0in]{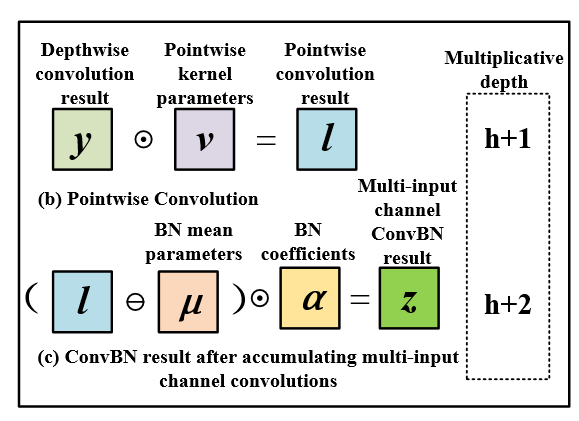}
			\label{Fig:sub1}
		\end{minipage}
	}
	\vspace{0.01cm}
	\subfigure[ConvBN with multiplicative-depth optimization]{
		\begin{minipage}[t]{0.5\textwidth}
			\centering
			\includegraphics[width=3.4in, height=2.0in]{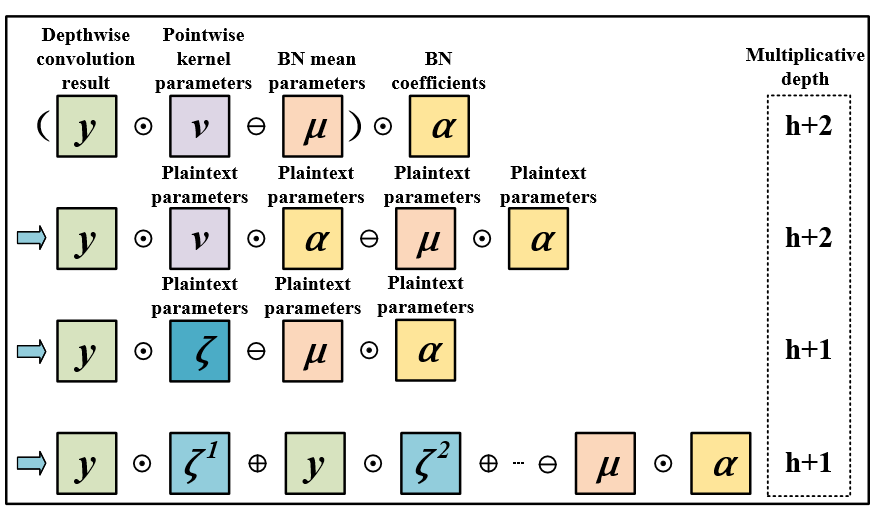}
			\label{Fig:sub2} 
		\end{minipage}
	}
	\caption{Schematic diagram of ConvBN.}
	\label{fig:ConvBNScheme}
\end{figure*}

Fig. \ref{fig:ConvBNScheme} sketches the diagram of our proposed ConvBN with/without optimization on multiplicative depth consumption. For the optimization, the coefficient 
\begin{math}
	\zeta
\end{math}
denotes the BN dot-product fusion matrix aggregated from the pointwise kernel parameters 
\begin{math}
	\nu
\end{math}
and the BN coefficients 
\begin{math}
	\alpha
\end{math}.
\begin{math}
	\zeta ^{n} 
\end{math}
is the rotated and filtered version of the BN dot-product fusion matrix. After the depthwise convolution result 
\begin{math}
	y 
\end{math}
is rotated, it is multiplied by
\begin{math}
	\zeta ^{n} 
\end{math},
and the products are accumulated, then, subtract
\begin{math}
	\mu \times \alpha 
\end{math}
to complete BN layer computation.

\section{Experiment Configuration} \label{Sec:Configuration}
\textbf{Running Environment.} The server is configured with two Intel Xeon Gold 6430 processors (Base Frequency 2.1 GHz/Max Turbo Frequency 3.4 GHz, 32 cores, 64 threads, 60 MB cache, 270WTDP), 8X 64GB Samsung DDRS ECC REGS 4800 memory modules, Ubuntu 22.04.

\textbf{Dataset and Models.} The dataset CIFAR-10 has 10 categories with each comprising 5,000 training images and 1,000 testing images. This dataset is widely used by convolutional neural networks in computer vision tasks, e.g. object recognition, despite the dataset is small-scale, it does not influence our evaluation, this is because our method focuses on the metrics of efficiency and accuracy loss over FHE. Therefore, we can employ the pretrained models in the plaintext environment, such as ResNet20/32/44. The ResNet20's rebuilt architecture is shown in Fig. \ref{fig:ResNetArchiFHE}, analogously, the ResNet32/44 can be reshaped as well.

\textbf{Encryption-Parameter Settings.} Regarding the encryption parameters towards the reshaped architectures of ResNet20/32/44, we set the ring dimension to $D=2^{16}$, the number of slots to $n=2^{14}$, the secret key's Hamming weight to $	h=64$, and the multiplicative depth to 26, with bootstrapping consuming 14 layers of multiplicative depth. The security level is set to $\lambda \ge 128$.

\textbf{Baselines.}  Since the previous PPML models, such as CryptoNet \cite{gilad2016cryptonets}, SEALion \cite{TimvanElsloo2019}, CryptoDL \cite{hesamifard2018privacy}, were applied into simple machine learning models with small number of layers, or replaced the non-arithmetic activation with simple arithmetic activation disregarding the usage of highly-precise polynomial approximation accompanied with bootstrapping operations. Hence, it is unsuitable to compare with them. We in this paper refer to the most related methods MPCNN\cite{lee2022low}, AESPA\cite{park2022aespa} and AutoFHE\cite{ao2024autofhe} as the baselines as shown in Table \ref{tab:Statistics}, wherein the brief statistics information is exhibited. 
To our best knowledge, the related work is limited to date.

\begin{table}[htbp]
	\centering
	\caption{Statistics on baselines and our FastFHE.}
	\label{tab:Statistics}
	\begin{tabular}{>{\centering\arraybackslash}p{1.7cm} >{\centering\arraybackslash}p{0.95cm} >{\centering\arraybackslash}p{1.25cm} >{\centering\arraybackslash}p{1.45cm} >{\centering\arraybackslash}p{1.25cm}}
		\toprule
		\textbf{Model Name} & \textbf{Venue} & \textbf{Encryption} & \textbf{Polynomial} & \textbf{Activation} \\
		\midrule
		MPCNN\cite{lee2022low} & ICML'22 & CKKS & high-degree &ReLU  \\
		AESPA\cite{park2022aespa} & arXiv'22 & CKKS & low-degree & ReLU  \\
		AutoFHE\cite{ao2024autofhe} & USENIX'24 & CKKS & multi-degree & mixed  \\
		FastFHE & ---- & CKKS & low-degree & SiLU \\
		\bottomrule
	\end{tabular}
\end{table}

MPCNN\cite{lee2022low}, AESPA\cite{park2022aespa} and AutoFHE\cite{ao2024autofhe} all adopt multi-channel packing schemes along with traditional convolution algorithms. Specifically, MPCNN focuses on the implementation of standard convolution under a single-ciphertext multi-channel setting and leverages high-degree polynomial approximations to improve the expressiveness of the ReLU function, thereby enhancing the inference accuracy over FHE. However, its high computational complexity results in prolonged inference time.
AESPA utilizes low-degree Hermite polynomials to approximate the ReLU activation function, remarkably reducing the computational cost over FHE, as well as decreasing the number of bootstrapping operations and overall inference latency, nevertheless, its poor approximation on activation function significantly degrades inference accuracy.
AutoFHE designs a selection algorithm that adaptively chooses polynomials of different degrees to approximate the activation function at different network layers. However, it cannot address the high computation cost problem. In their framework, achieving high accuracy still requires a considerable inference time.

\section{Related Work} \label{Sec:relatedWork}
\subsection{Computation-Load Reduction of Linear Layer}

With the advancement of fully homomorphic encryption (FHE) techniques in recent years, Cheon et al. \cite{cheon2018bootstrapping} proposed a bootstrapping algorithm under the RNS-CKKS scheme, which made it feasible to perform privacy-preserving machine learning (PPML) over FHE. Since then, researchers have turned their attention to implementing PPML based on RNS-CKKS. However, turning this concept into a practical solution remains challenging. Lee et al. \cite{lee2022privacy} were the first to realize privacy-preserving deep neural network inference using bootstrapping, but their work did not consider the architectural constraints imposed by the FHE environment. As a result, their model suffered from long inference times. To address this, researchers begin focusing on the computational cost of linear layers, which remains one of the primary bottlenecks in FHE-based deep learning inference. Juvekar et al. \cite{juvekar2018gazelle} propose GAZELLE, a framework that combines secure multi-party computation and homomorphic encryption to enable efficient neural network inference. They introduce an effective convolution method under homomorphic encryption that significantly reduces the number of homomorphic operations for standard convolutions. However, in deeper networks, as the numbers of input and output channels grows, the convolution of multiple ciphertexts still incurs high time and memory costs. Building on this, Lee et al. \cite{lee2022low} propose a more efficient convolution scheme and successfully implement deep neural networks over FHE. Nonetheless, due to the inherently high computation complexity of traditional convolutions and the need to enlarge encryption parameters for packing multi-channel ciphertext data, their method still requires prolonged inference time and consume substantial memory resources. Rovida et al. \cite{rovida2024encrypted} introduce an optimized vector-encoded convolution method that requires fewer rotation keys. However, due to the relatively outdated packing strategies and ciphertext slot utilization, the number of rotation operations increase significantly. Although their method reduces the memory consumption, the excessive homomorphic rotations lead to long computation times for linear layers and negatively impacts the overall model performance.

Despite significant progress has been made in ciphertext-domain convolution, existing approaches still incur substantial computational overhead when performing convolutions on encrypted data. The time-consuming overhead hinders the practical deployment of privacy-preserving inference over FHE. In this regard, our FastFHE addresses the challenge by employing a scalable data-packing scheme for ciphertext, which compactly encodes multi-channel data into a single ciphertext. This strategy significantly reduces the number of homomorphic rotation operations required during linear layer computation, thereby lowering memory consumption. Additionally, our proposed FastFHE introduces a specially designed ConvBN architecture tailored to this packing scheme, further reducing the time cost of ciphertext convolutions. Together, these innovations effectively mitigate the excessive computational burden of linear layers.

\subsection{Approximate Expression of Activation Layer}

Scholars have found that, in order to maintain model inference accuracy over FHE, the use of bootstrapping operations must be carefully balanced. Too few bootstrapping operations can result in excessive noise accumulation, leading to significant inference error or decryption failure. Conversely, excessive bootstrapping significantly increases the overall inference time, making it another major bottleneck in FHE-based model inference. For FHE, the number of bootstrapping operations required is primarily determined by the multiplicative depth supported by the encryption parameters. Currently, the majority of multiplicative depth is consumed by non-linear activation functions, since the addition and multiplication alone cannot express non-linear transformations. A few researchers, such as Crawford et al. \cite{crawford2018doing}, have attempted to approximate activation functions using lookup tables for low-precision estimation. However, this often leads to considerable accuracy degradation, which prompts to adopt polynomial approximation technique instead. Gilad-Bachrach et al. \cite{gilad2016cryptonets} propose a square function to substitute the activation function, which yields satisfied results in a shallow five-layer network. However, they also point out several issues associated with square activation, such as gradient explosion during training, limited expressiveness, and hindered learning capacity. These problems would become more severe in deeper networks. Brutzkus et al. \cite{brutzkus2019low} also employ square activation in their LoLa system. Bourse et al. \cite{bourse2018fast} propose an alternative approach by developing a method to compute nonlinear activation using sign functions. Lee et al. \cite{lee2022low} empirically select a composite polynomial composed of minimax approximations of degree-5/15/27, which offers small approximation errors. However, this method consumes a large amount of multiplicative depth, requiring frequent bootstrapping operations and leading to an inference time of approximately 3 hours under the environment: AMD Ryzen Threadripper PRO 3995WX at 2.096 GHz (64 cores) with 512 GB RAM. Park et al. \cite{park2022aespa} adopt a degree-2 Hermite polynomial to approximate the ReLU function. This design significantly reduces bootstrapping overhead due to its low multiplicative depth consumption. Nonetheless, the coarse-fined approximation results in a substantial drop in inference accuracy. Resorting to modeling the time overhead problem into a multi-objective optimization framework, Ao et al. \cite{ao2024autofhe} recently propose hierarchical mixed-degree polynomial activation to maximize accuracy while minimizing bootstrapping operations. However, their inference framework still faces a trade-off problem between accuracy and inference time, and the complex network architecture demands a high-performance deployment environment (at least 768 GB of RAM).

Towards this dilemma, our work proposes a novel activation function approximation strategy, aiming to optimize multiplicative depth consumption while improving the approximation quality of nonlinear activation functions. Specifically, a low-degree Legendre polynomial is used to approximate the smooth function SiLU, which has better representational capabilities. This approximation achieves a high level of inference accuracy while consuming only three levels of multiplicative depth. As a result, it significantly enhances the expressiveness of activation layers in the ciphertext domain, allowing the use of pretrained model parameters from plaintext environments. Our approach improves the overall model inference accuracy and effectively addresses the issue of excessive multiplicative depth consumption and poor performance on the approximation quality of existing low-degree polynomials.

\end{document}